\documentclass[aps,prb,twocolumn,superscriptaddress,nofootinbib]{revtex4-2}

\usepackage{graphicx}
\usepackage{float}
\usepackage{amsmath,amssymb,bm}
\usepackage{dcolumn}
\usepackage{color}
\usepackage{hyperref}

\hypersetup{
  colorlinks=true,
  citecolor=blue,
  linkcolor=blue,
  urlcolor=blue
}

\newcommand{\tp}{t_{\perp}}
\newcommand{\tpc}{t_{\perp,c}}
\newcommand{\dd}{\delta}
\newcommand{\Ga}{\Gamma\text{-}\alpha}
\newcommand{\Mb}{\mathrm{M}\text{-}\beta}
\newcommand{\Sn}{\Sigma^{\rm nor}}
\newcommand{\Sa}{\Sigma^{\rm ano}}

\newcommand{\safeincludegraphics}[2][]{%
  \IfFileExists{#2}{\includegraphics[#1]{#2}}{%
    \fbox{\begin{minipage}[c][0.16\textheight][c]{0.88\linewidth}
    \centering Missing figure file: \texttt{\detokenize{#2}}
    \end{minipage}}%
  }%
}

\begin{document}

\title{Self-energy pole optimization of superconductivity in the bilayer Hubbard model}

\author{Taka-Shi Fujiwara}
\affiliation{Department of Physics, The University of Tokyo, 7-3-1 Hongo, Bunkyo-ku, Tokyo 113-0033, Japan}

\author{Shiro Sakai}
\affiliation{Physics Division, Sophia University, Chiyoda-ku, Tokyo 102-8554, Japan}

\author{Ryotaro Arita}
\affiliation{Department of Physics, The University of Tokyo, 7-3-1 Hongo, Bunkyo-ku, Tokyo 113-0033, Japan}
\affiliation{RIKEN Center for Emergent Matter Science, 2-1 Hirosawa, Wako, Saitama 351-0198, Japan}

\date{\today}

\begin{abstract}
We 
study the real-frequency structure of the self-energy in the bilayer Hubbard model,
using the dynamical cluster approximation.
At half filling, the Mott insulator--band insulator (MI--BI) crossover involves a rearrangement of self-energy poles between the bonding and antibonding bands; these poles cross as the interlayer hopping $\tp$ is varied.
 Upon doping, this pole structure produces a 
band-selective pseudogap and enhances $s^{\pm}$-wave superconductivity.
The order parameter is maximized near the MI--BI boundary, where low-energy anomalous self-energy poles develop simultaneously in both 
bands
and cooperatively enhance the pairing. We further show that 
these self-energy poles can be interpreted as 
emergent fermionic excitations, offering an enhanced-pairing mechanism in common 
with the single-layer Hubbard model.
The controllability of these poles through $\tp$ 
makes the bilayer system an unconventional platform for optimizing strongly correlated superconductivity.
\end{abstract}

\maketitle

\textit{Introduction.}
In strongly correlated electron systems, the electron self-energy can have a nontrivial frequency dependence beyond a mere renormalization of the quasiparticle mass: its low-energy poles reshape the single-particle excitation spectrum, producing features such as the Mott gap and pseudogap. 
When superconductivity occurs due to the strong correlations,
the dynamical structure of the anomalous self-energy---not just its static value---controls the pairing strength \cite{Maier2008,Kyung2009,Civelli2009,GullMillis2015,Sakai2016PRL}, signaling the underlying unconventional mechanism, just as 
in conventional superconductors, where the analysis of the frequency dependence identified their phonon-mediated mechanism~\cite{Eliashberg1960,McMillanRowell1965,Scalapino1966}.

Such physics is most pronounced near a Mott insulator, where strong correlations generate a large self-energy with marked momentum and frequency dependence---a picture widely invoked for the pseudogap and superconductivity in underdoped cuprates \cite{Anderson1987,Keimer2015,LeeNagaosaWen2006,Scalapino2012}. Indeed, cluster dynamical mean-field studies of the single-layer Hubbard model have found that self-energy poles inherited from the Mott insulator govern the low-energy dynamics of both the pseudogap and the superconducting states
\cite{Stanescu2006,Sakai2009,Gull2013,Civelli2009,Sordi2012,GullMillis2015,Sakai2016PRL,Sakai2016PRB,Sakai2018PRB}.
 In particular, 
 the poles of the anomalous self-energy
 enhance the pairing 
 \cite{Maier2008,Sakai2016PRL}. In the single-layer model, however, the pole position is 
 mostly determined
 by the 
 interaction strength
 and the doping level \cite{Sakai2016PRB}, leaving little room for 
 tuning it 
 to strengthen superconductivity.

Recently, the bilayer Hubbard model has attracted renewed attention as another
simple model accommodating strongly correlated superconductivity
\cite{Bulut1992,Scalettar1994,Hetzel1994,DosSantos1995,Liechtenstein1995,%
Bouadim2008,Lanata2009,Maier2011,Maier2019,Karakuzu2021},
especially in light of the pressure-induced high-$T_c$ superconductivity in the
bilayer nickelate La$_3$Ni$_2$O$_7$~\cite{Sun2023} and the subsequent experimental \cite{Hou2023,Wang2024PRX,Zhang2024NatPhys,Wang2024Pr,Ko2025,Zhou2025,Liu2025NatMater,Hao2025NatMater,Li2026Sm}
and 
theoretical studies
\cite{Luo2023,Christiansson2023,Lechermann2023,Liu2023,Yang2023,Liao2023,Zhang2024Nickelate,Heier2024,Botzel2024,Ryee2024,Jiang2024CPL,Luo2024NPJ,Lu2024,Jiang2025PRL,Tian2024,Fan2024,Gu2025,Sakakibara2024}.
In this model, the interlayer hopping $\tp$ splits the spectrum into bonding and antibonding bands and controls unconventional pairing channels.
The model is known to host a Mott insulator--band insulator (MI--BI) crossover at half filling \cite{KancharlaOkamoto2007,Lee2014}, as well as pseudogap behavior \cite{Fratino2022,Karakuzu2021,Nomura2025} and $s^{\pm}$-wave superconductivity with a rather high critical temperature \cite{Kuroki2002,Mishra2016,Kato2020,Yang2023,Liu2023,Botzel2024,Heier2024,
Sakakibara2024,Karakuzu2021,Nomura2025} upon light hole doping.
However, its superconducting mechanism has been mainly discussed in terms of
weak-coupling theory, where the shape of the Fermi surfaces (i.e., poles of the
Green's function) is a key parameter
\cite{Kuroki2002,Mishra2016,Yang2023,Liu2023,Botzel2024,Heier2024,%
Sakakibara2024}.

In this work, we demonstrate a Mott-physics-driven high-temperature superconducting mechanism, which is controllable through $\tp$, in the small doping region of the bilayer Hubbard model.
We compute the real-frequency self-energy, 
using the dynamical cluster approximation \cite{Hettler1998,Maier2005RMP} combined with exact diagonalization (DCA+ED). 
At half filling, the MI--BI crossover \cite{KancharlaOkamoto2007,Lee2014} appears 
as a spectral reconstruction accompanied by 
a rearrangement of self-energy poles between the 
bonding and antibonding bands.
These poles move through the low-energy region and cross each other as $\tp$ is varied. Upon doping, this structure develops into 
band-selective pseudogaps in the normal state and enhances an $s^{\pm}$-wave superconducting state. In the latter case, the order parameter is maximized near the MI--BI boundary, where the self-energy poles of both bands simultaneously reach low energy and cooperatively enhance 
the superconductivity.
We show that the 
low-energy structure of the self-energy is
quantitatively reproduced by a hidden-fermion model \cite{Sakai2016PRL,Sakai2016PRB,Sakai2023}  extended to the bilayer system, identifying the low-energy poles as emergent fermionic excitations, common to those in the single-layer 
Hubbard 
model.
The controllability of the self-energy poles through $t_\perp$ in the bilayer model provides
a microscopic handle for designing correlated superconductivity.
\begin{figure*}[t]
\centering
\safeincludegraphics[width=\linewidth]{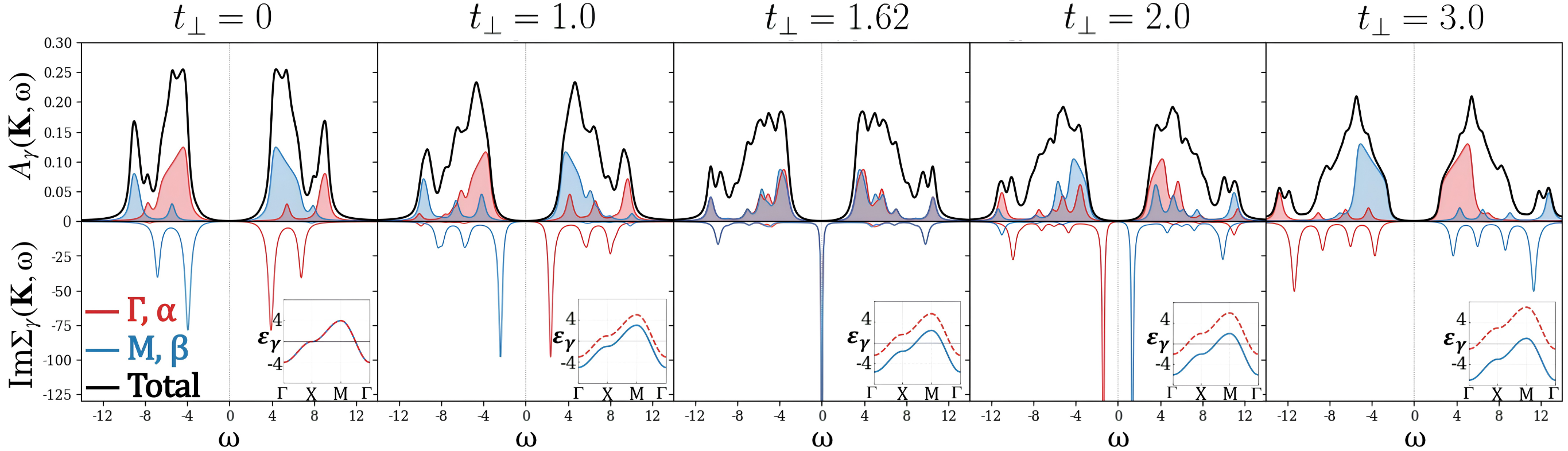}
\caption{Spectral function $A_{\gamma}(\mathbf{K},\omega)$ (top row) and imaginary part of
the normal self-energy ${\rm Im}\,\Sn_{\gamma}(\mathbf{K},\omega)$ (bottom row) at half
filling ($\dd=0$) and $U=12$ for $\tp=0$, $1.0$, $1.62$, $2.0$, and $3.0$ (left to
right). Red and blue curves denote the $\Ga$ and $\Mb$ components; the black
curve in the top row is the total DOS. The poles of ${\rm Im}\,\Sn_{\gamma}$, located
inside the Mott gap, cross around $\tp\simeq\tpc\simeq 1.62$, identifying the MI--BI
crossover as a rearrangement of these poles between the two components. Insets in
the lower panels show the bare ($U=0$) band dispersions
$\varepsilon_{\gamma}(\mathbf{k})$ along $\Gamma$-$\mathrm{X}$-$\mathrm{M}$-$\Gamma$.}
\label{fig:half_filling}
\end{figure*}

\textit{Model and method.} 
We consider the bilayer Hubbard model on a square lattice, 
which reads
\begin{equation}
\resizebox{0.89\columnwidth}{!}{$\displaystyle
\mathcal{H}
=
-t\sum_{\langle i,j\rangle,l,\sigma}
c^{\dagger}_{il\sigma}c_{jl\sigma}
-\tp\sum_{i,l\neq l',\sigma}
c^{\dagger}_{il\sigma}c_{il'\sigma}
+U\sum_{i,l}
n_{il\uparrow}n_{il\downarrow}
$}.
\label{eq:bilayer_hubbard}
\end{equation}
Here, $i$ and $j$ label lattice sites, $l,l'=1,2$ 
are layers,
and $\sigma=\uparrow,\downarrow$ 
are spins.
The parameters $t$ ($\tp$) and $U$ denote the intralayer (interlayer) nearest-neighbor hopping and the onsite Coulomb repulsion, respectively. We set $t=1$ as the energy unit and take $\tp>0$ throughout the paper. 
The 
chemical potential $\mu$ 
controls
the hole-doping concentration 
$\dd$ w.r.t. half filling.

Introducing the bonding--antibonding basis 
between the layers,
$c^{\alpha}_{i\sigma}=\frac{1}{\sqrt{2}}(c_{i1\sigma}-c_{i2\sigma})$ and $c^{\beta}_{i\sigma}=\frac{1}{\sqrt{2}}(c_{i1\sigma}+c_{i2\sigma})$, we obtain the band dispersions 
$\varepsilon_{\alpha,\beta}(\mathbf{k})=-2t(\cos k_x+\cos k_y)\pm\tp -\mu$ for $U=0$.
Hereafter, we specify
the band 
with
$\gamma\in\{\alpha,\beta\}$, where $\alpha$ and $\beta$ denote the antibonding and bonding bands, respectively.

To treat short-range correlations and symmetry breaking, we employ the dynamical cluster approximation (DCA) 
\cite{Hettler1998,Maier2005RMP}. In DCA, the Brillouin zone is divided into patches centered at cluster momenta $\mathbf{K}$, and the lattice problem is mapped onto a finite cluster impurity problem through the coarse-grained Green's function
\begin{align}
\bar{G}_{\gamma}(\mathbf{K},\omega)
=
\frac{N_c}{2N}
\sum_{\tilde{\mathbf{k}}}
G_{\gamma}(\mathbf{K}+\tilde{\mathbf{k}},\omega).
\label{eq:dca_coarse}
\end{align}
Here, $N$ denotes the number of lattice sites in each layer, $N_c$ is the number of sites in a cluster, and $\tilde{\mathbf{k}}$ runs over the $2N/N_c$ momentum points 
in each patch.
Besides the layers, we
take two cluster momenta $\mathbf{K}=(0,0)$ and $(\pi,\pi)$ ($N_c=4$ in total), 
around which
separate Fermi surfaces formed by the $\alpha$ and $\beta$ bands reside for $U=0$ and $0<\tp<4$. We label these low-energy components 
$\Ga \equiv (\mathbf{k}=(0,0),\alpha)$ and
$\Mb \equiv (\mathbf{k}=(\pi,\pi),\beta)$. 

The four-site interacting cluster problem with
eight noninteracting bath sites is solved by the Lanczos method \cite{Caffarel1994}, which can yield real-frequency Green's functions directly via continued-fraction expansions \cite{Gagliano1987}, without analytic continuation from the Matsubara axis. 
The bath
parameters are determined self-consistently, with an artificial 
temperature 
$T=t/200$,
which 
is low enough to effectively yield
the ground state.

In the superconducting state, 
the normal Green's function is written as
\begin{align}
\resizebox{0.9\columnwidth}{!}{$\displaystyle
G_{\gamma}(\mathbf{k},\omega)
=
\left[
\omega-\varepsilon_{\gamma}(\mathbf{k})
-\Sn_{\gamma}(\mathbf{k},\omega)
-W_{\gamma}(\mathbf{k},\omega)
\right]^{-1},$}
\label{eq:G_with_W}
\end{align}
where $\Sn_{\gamma}(\mathbf{k},\omega)$ is the normal self-energy and
\begin{align}
\resizebox{0.9\columnwidth}{!}{$\displaystyle
W_{\gamma}(\mathbf{k},\omega)
=
\left[\Sa_{\gamma}(\mathbf{k},\omega)\right]^2
\left/(
\omega+\varepsilon_{\gamma}(\mathbf{k})
+\Sn_{\gamma}(\mathbf{k},-\omega)^{*}
\right)$}
\label{eq:w_def}
\end{align}
represents the feedback of the anomalous self-energy $\Sa_{\gamma}(\mathbf{k},\omega)$ on 
$G_{\gamma}(\mathbf{k},\omega)$.
The superconducting gap appearing in the spectral function $A_{\gamma}(\mathbf{K},\omega)=-(1/\pi){\rm Im}\bar{G}_{\gamma}(\mathbf{K},\omega)$ is thus determined by the 
total
self-energy $\Sn_{\gamma}+W_{\gamma}$. 
We substitute the 
self-energy, $\Sn_{\gamma}(\mathbf{K},\omega)$ and $\Sa_{\gamma}(\mathbf{K},\omega)$, at cluster momenta into Eqs.~(\ref{eq:G_with_W}) and (\ref{eq:w_def}) 
to obtain $\bar{G}_{\gamma\sigma}$ and $W_{\gamma}$.

\textit{Spectra at half filling.} 
Figure~\ref{fig:half_filling} shows the spectral functions and normal self-energies at half filling ($\dd=0$) and $U=12$ for various values of $\tp$. At $\tp=0$, the two layers are decoupled, and the system 
reduces to the single-layer Hubbard model in the Mott-insulating regime. In this case, the spectral weight around the Fermi level is strongly suppressed
due to the poles of $\Sn_{\gamma}$, which appear as sharp peaks of ${\rm Im}\,\Sn_{\gamma}$ in our results.
As the self-energy pole of $\Ga$ ($\Mb$) is located on the unoccupied (occupied) side, the spectral weight of $\Ga$ ($\Mb$) is pronounced on the occupied (unoccupied) side.

As $\tp$ increases, the spectral center of 
$\Ga$ ($\Mb$) shifts upward (downward)
owing to the increasing bonding--antibonding splitting. Concomitantly, the self-energy pole of $\Ga$ ($\Mb$) inside the gap 
shifts downward (upward),
crossing each other at $\tp\simeq\tpc$, where $\tpc\simeq 1.62$ 
for $U=12$. For $\tp>\tpc$, they retreat from low energy, and the spectra become band-insulator-like, dominated by the bonding--antibonding splitting. 
Thus, the
MI--BI crossover at half filling \cite{KancharlaOkamoto2007,Lee2014}
does
not 
involve a gap closing
but a 
spectral
rearrangement 
between the $\Ga$ and $\Mb$ components. Crucially, the controllability of these self-energy poles by $\tp$, uncovered here at half filling, carries over to the doped region 
and 
governs the pseudogap and superconducting states there, as we shall see in the following.

\textit{Pseudogap state.} 
At small doping ($\dd=0.02$), we find band-selective pseudogaps by constraining the solution to the normal state. 
The 
spectra at $U=12$ are shown in Sec.~S2 of the Supplemental Material. For $\tp=1\lesssim\tpc$, 
the $\Mb$ ($\Ga$) component is gapped (gapless) \cite{Fratino2022}. 
For $\tp=2\gtrsim\tpc$,
the gapped (gapless) component switches to $\Ga$ ($\Mb$) \cite{Karakuzu2021,Nomura2025}. Thus, in the lightly hole-doped regime, two types of band-selective pseudogaps emerge, reflecting the self-energy poles that remain 
on
the occupied side at half filling: an $\Mb$ ($\Ga$)-selective one for $\tp\lesssim\tpc$ ($\tp\gtrsim\tpc$). This component selectivity is characteristic 
of the bilayer system and is inherited by the anomalous self-energy in the superconducting state.
\begin{figure}[t]
\centering
\safeincludegraphics[width=1.02\linewidth]{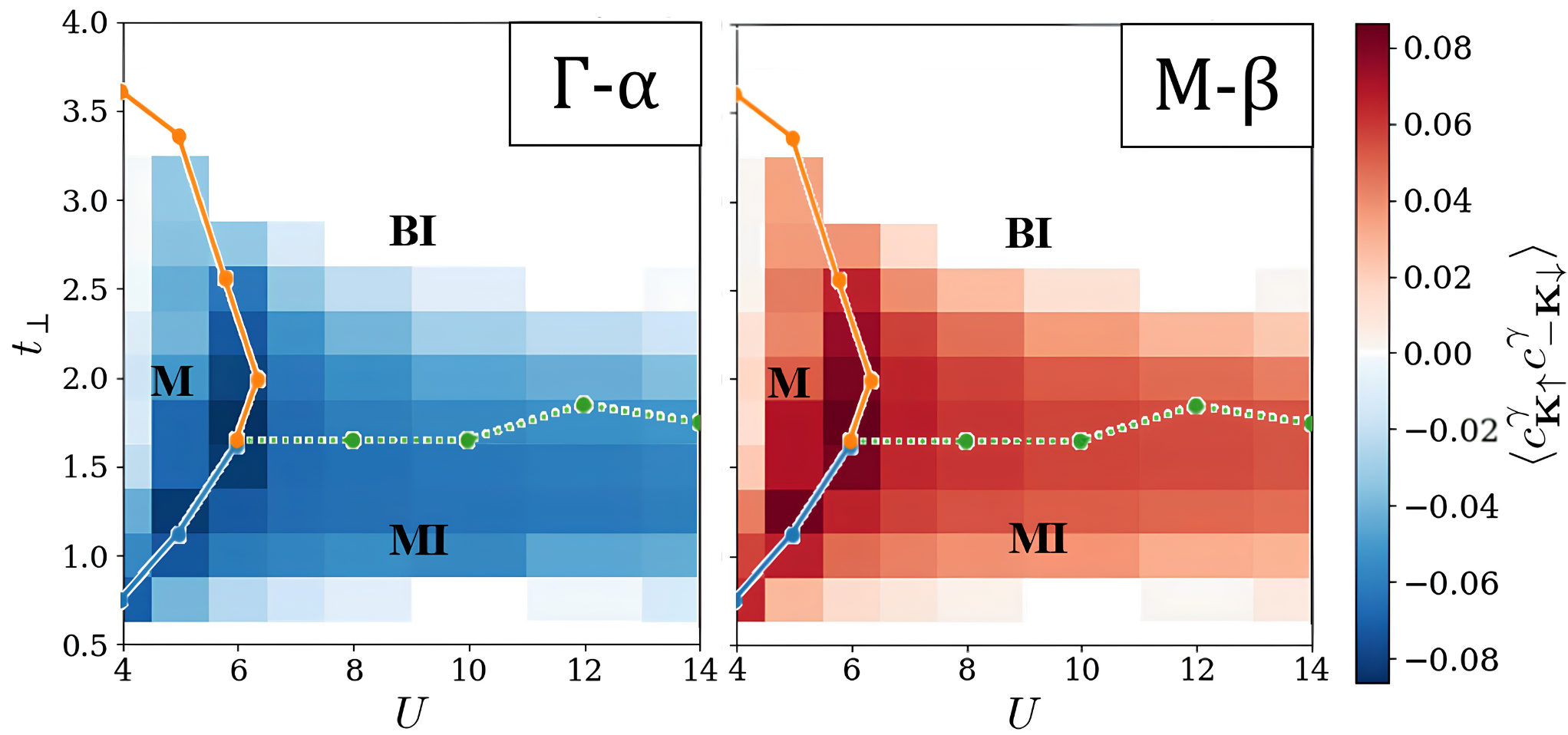}
\caption{Superconducting order parameters in the $(U,\tp)$ plane at $\dd=0.02$:
$\langle c_{\Gamma\uparrow}^{\alpha}c_{\Gamma\downarrow}^{\alpha}\rangle$ for the
$\Ga$ component (left) and
$\langle c_{\mathrm{M}\uparrow}^{\beta}c_{\mathrm{M}\downarrow}^{\beta}\rangle$ for the $\Mb$
component (right), on a common color scale. The opposite signs ($\Ga<0$: blue,
$\Mb>0$: red) indicate $s^{\pm}$-wave symmetry. The 
phase diagram at half filling is
superimposed (M: metal, MI: Mott insulator, BI: band insulator; the MI--BI boundary is
determined in Sec.~S5). Both components are largest near the MI--BI boundary and
peak toward the metallic region.}
\label{fig:op_map}
\end{figure}

\textit{Superconducting order parameter.} 
Figure~\ref{fig:op_map} shows the $(U,\tp)$ 
map
of the order parameters at $\dd=0.02$ 
for the $\Ga$ and $\Mb$ components, with the half-filling phase diagram superimposed.
The two components have opposite signs and approximately satisfy
$\langle c_{\Gamma\uparrow}^{\alpha}c_{\Gamma\downarrow}^{\alpha}\rangle
\simeq
-\langle c_{\mathrm{M}\uparrow}^{\beta}c_{\mathrm{M}\downarrow}^{\beta}\rangle$,
consistent with the
$s^{\pm}$-wave symmetry, with a sign change between the $\Ga$ and $\Mb$ components \cite{Karakuzu2021,Nomura2025}.

This $s^{\pm}$-wave superconductivity emerges mainly in the $(U,\tp)$ region
where the self-energy poles generating the Mott gap reside at low energy for
both components at half filling (see Fig.~\ref{fig:half_filling}).
Accordingly, the magnitude of the order parameter becomes large for both components near 
the MI--BI boundary, where these two poles approach and cross each other at low energy. This enhancement is therefore tied to the 
self-energy poles present at half filling and 
persisting at low energy even after doping. Moreover, along the MI--BI boundary, the order parameter is maximized near 
the boundary with the metallic region at half filling. This is likely because the poles of 
both
components, originally located inside a narrow gap at half filling, remain very close to the Fermi level upon doping and efficiently enhance pair formation, as 
we shall discuss
in the 
following.

\textit{Real-frequency spectra in the superconducting state.} 
\begin{figure}[t]
\centering
\safeincludegraphics[width=\linewidth]{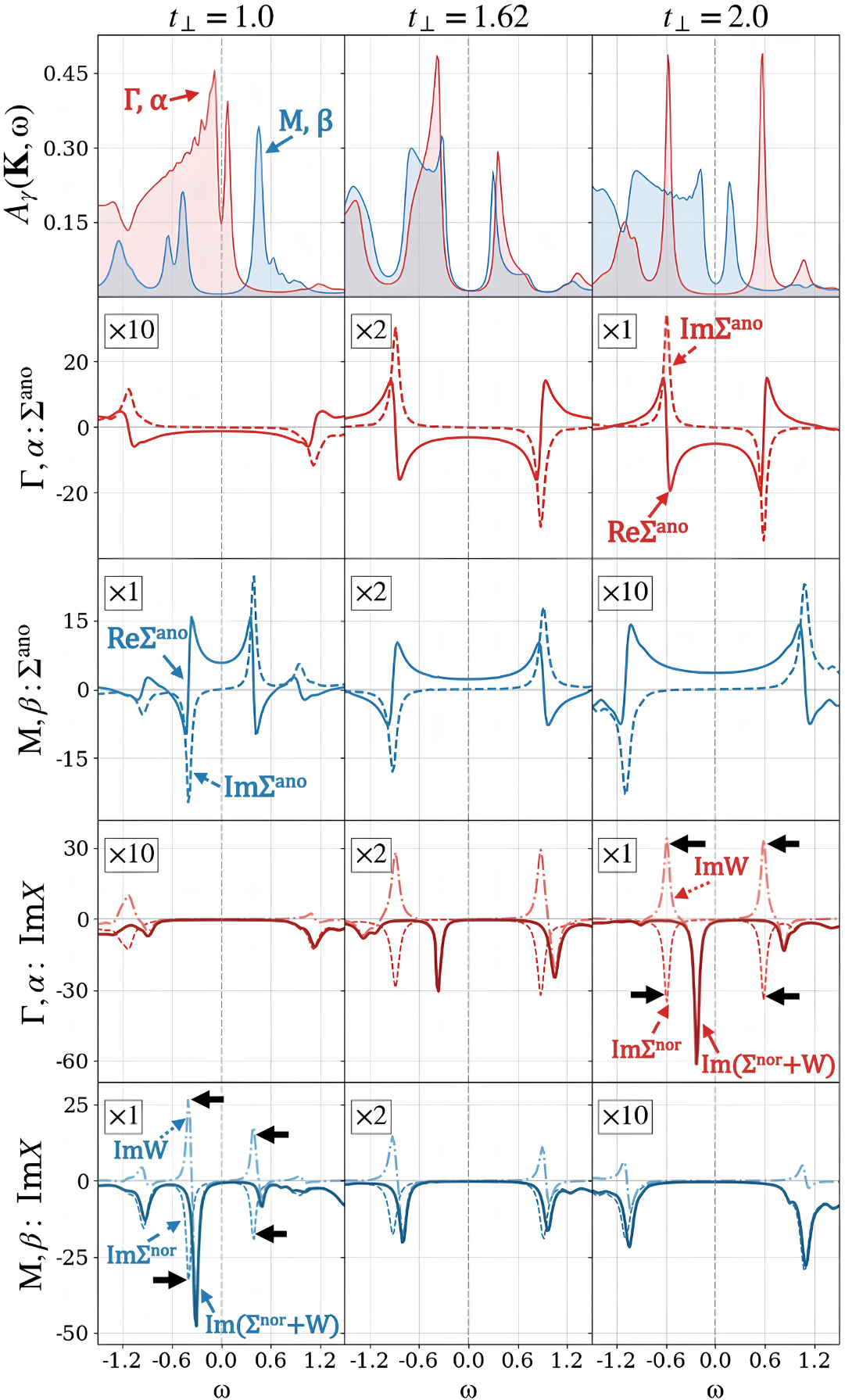}
\caption{Real-frequency results in the superconducting state at $U=12$ and
$\dd=0.02$ for $\tp=1.0$, $1.62\,(\simeq\tpc)$, and $2.0$ (left to right). 
From top to bottom: spectral function
$A_{\gamma}(\mathbf{K},\omega)$ ($\Ga$: red, $\Mb$: blue); anomalous self-energy
$\Sa_{\gamma}(\mathbf{K},\omega)$ for $\Ga$ and then $\Mb$ (solid: ${\rm Re}\,\Sa$, dashed:
${\rm Im}\,\Sa$); and ${\rm Im}\,X_{\gamma}$ for $\Ga$ and then $\Mb$ with
$X\in\{\Sn,W,\Sn+W\}$ (${\rm Im}\,\Sn_{\gamma}$: dashed, ${\rm Im}\,W_{\gamma}$:
dash-dotted, sum: solid). Per-panel scale factors (except $A_{\gamma}$) are given
in the upper-left corner. Black arrows mark the pole-to-pole cancellation between
$\Sn_{\gamma}$ and $W_{\gamma}$, derived analytically in Sec.~S3.}
\label{fig:sc_spectra}
\end{figure}
Figure~\ref{fig:sc_spectra} shows 
$A_{\gamma}$, $\Sa_{\gamma}$,
and ${\rm Im}\,W_{\gamma}$ in the superconducting state at $U=12$ and $\dd=0.02$. We compare the representative values $\tp=1.0$, $1.62\,(\simeq\tpc)$, and $2.0$, which straddle the pole crossing 
at half filling. We find that a superconducting gap opens in both the $\Ga$ and $\Mb$ components for all these values of $\tp$, and that anomalous self-energy poles stand out at particle-hole symmetric energies in both components.

As in the pseudogap state, the structure of the self-energy poles at half filling (Fig.~\ref{fig:half_filling}) is carried over into the superconducting state for $\delta>0$, not only in the normal but also in the anomalous component.
For $\tp=1<\tpc$, 
the $\Mb$ pole dominates the low-energy region after doping; for $\tp=2>\tpc$,
the $\Ga$ pole dominates instead.
At $\tp\simeq\tpc$, 
both poles are around zero energy at half filling. This generates pronounced low-energy 
poles in $\Sa_\gamma$ for both $\Ga$ and $\Mb$ after hole doping, though the poles slightly shift away from the Fermi energy due to the hole doping.

These poles strongly enhance the static pairing 
strength
${\rm Re}\,\Sa_{\gamma}(\mathbf{K},\omega=0)$ through the Kramers--Kronig relation
\begin{align}
{\rm Re}\,\Sa_{\gamma}(\mathbf{K},\omega=0)=\frac{2}{\pi}\int_{0}^{\infty}d\omega\,
\frac{{\rm Im}\,\Sa_{\gamma}(\mathbf{K},\omega)}{\omega},
\label{eq:kk_anomalous}
\end{align}
and thereby enhance the superconducting gap \cite{Maier2008}. Since the integrand on the r.h.s. of Eq.~(\ref{eq:kk_anomalous}) is weighted by $1/\omega$, a pole closer to $\omega=0$ gives a larger contribution. At $\tp\simeq\tpc$, such low-energy poles exist simultaneously in both $\Ga$ and $\Mb$ components, and hence ${\rm Re}\,\Sa_{\gamma}(0)$ is enhanced for both. This directly produces the order-parameter maximum in Fig.~\ref{fig:op_map}: the cooperative growth of both superconducting gaps strengthens the $s^{\pm}$ order near the MI--BI boundary. 

As shown in the lower panels of Fig.~\ref{fig:sc_spectra}, a nontrivial cancellation occurs between ${\rm Im}\,\Sn_{\gamma}$ and ${\rm Im}\,W_{\gamma}$ in the superconducting state: Although both $\Sn_{\gamma}$ and $W_{\gamma}$ 
exhibit pole structures, 
a pole disappears
in their sum
and hence
in the Green's function. This pole-to-pole cancellation has also been observed in the single-layer Hubbard model \cite{Sakai2016PRL,Sakai2018PRB}, 
where it indicates
that the poles of the normal 
and anomalous self-energies share the same origin.
This cancellation 
does not occur for a self-energy peak generated by bosonic excitations and instead suggests a coupling to an emergent fermionic excitation \cite{Yamaji2011,Sakai2016PRL,ImadaSuzuki2019},
as discussed in the 
following;
see Sec.~S3 of the Supplemental Material for 
details.

\textit{Hidden-fermion fitting.} 
\begin{figure}[t]
{\centering
\safeincludegraphics[width=\linewidth]{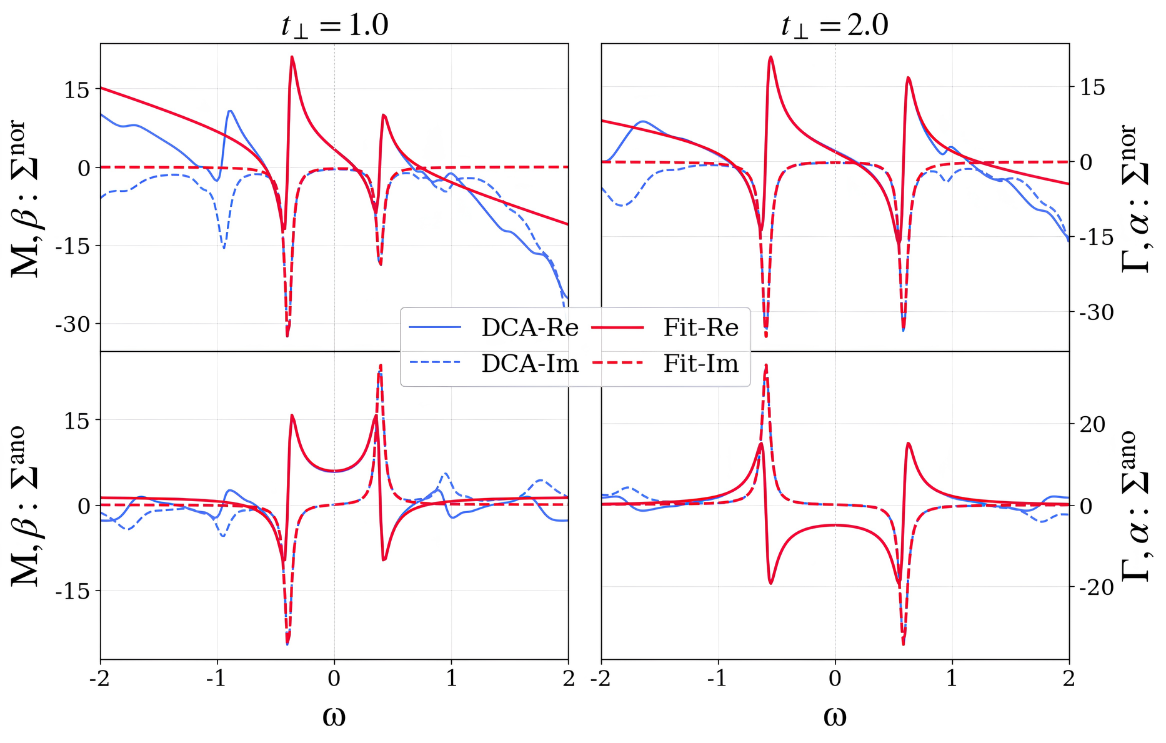}}
\caption{Hidden-fermion fit of 
$\Sn_{\gamma}(\mathbf{K},\omega)$ (top) and
$\Sa_{\gamma}(\mathbf{K},\omega)$ (bottom) 
at $U=12$ and
$\dd=0.02$: the $\Mb$ component at $\tp=1.0$ (left) and the $\Ga$
component at $\tp=2.0$ (right). Blue curves are the DCA
data and red
curves are the fitting with 
the hidden-fermion model [Eqs.~(\ref{eq:hf_normal}) and
(\ref{eq:hf_anomalous})]; solid and dashed 
curves
denote real and imaginary
parts. A single low-energy hidden fermion 
reproduces the pole structure
of both channels.}
\label{fig:hidden_fit}
\end{figure}
To interpret the DCA self-energies, we employ 
a minimal phenomenological model in which the physical electron $c$ hybridizes with an 
emergent
fermionic degree of freedom $f$ \cite{Sakai2016PRL,Sakai2016PRB,Sakai2023}:
\begin{align}
&\mathcal{H}^{\rm HF}
=
\sum_{\mathbf{k},\gamma,\sigma}
\left[
(\varepsilon_{c,\gamma}(\mathbf{k})+s_{\gamma\mathbf{k}}) c_{\mathbf{k}\sigma}^{\gamma\dagger}c_{\mathbf{k}\sigma}^{\gamma}
+
\varepsilon_{f,\gamma}(\mathbf{k}) f_{\mathbf{k}\sigma}^{\gamma\dagger}f_{\mathbf{k}\sigma}^{\gamma}
\right]
\nonumber\\
&+
\sum_{\mathbf{k},\gamma,\sigma}
V_{\gamma\mathbf{k}}
\left(
c_{\mathbf{k}\sigma}^{\gamma\dagger}f_{\mathbf{k}\sigma}^{\gamma}+{\rm H.c.}
\right)
\nonumber\\
&-
\sum_{\mathbf{k},\gamma}
\left(
D_{c,\gamma\mathbf{k}}
c_{\mathbf{k}\uparrow}^{\gamma\dagger}c_{\mathbf{k}\downarrow}^{\gamma\dagger}
+
D_{f,\gamma\mathbf{k}}
f_{\mathbf{k}\uparrow}^{\gamma\dagger}f_{\mathbf{k}\downarrow}^{\gamma\dagger}
+{\rm H.c.}
\right).
\label{eq:hidden_hamiltonian}
\end{align}
Here, $\varepsilon_{c,\gamma}$ and $\varepsilon_{f,\gamma}$ denote the dispersions of the $c$ and $f$ fermions, respectively, $s_{\gamma\mathbf{k}}$ is the energy shift emerging from the electron correlation, $V_{\gamma\mathbf{k}}$ is their hybridization, and $D_{c,\gamma\mathbf{k}}$ and $D_{f,\gamma\mathbf{k}}$ are their anomalous terms. Integrating out the $f$ degrees of freedom,
we obtain the self-energies for the $c$ fermion as
\begin{align}
\Sn_{\gamma}(\mathbf{k},\omega)
&=s_{\gamma\mathbf{k}}+
\frac{
V_{\gamma\mathbf{k}}^2(\omega+\varepsilon_{f,\gamma}(\mathbf{k}))
}{
\omega^2-\varepsilon_{f,\gamma}(\mathbf{k})^2-D_{f,\gamma\mathbf{k}}^2
},
\label{eq:hf_normal}
\\
\Sa_{\gamma}(\mathbf{k},\omega)
&=
D_{c,\gamma\mathbf{k}}
-
\frac{
V_{\gamma\mathbf{k}}^2 D_{f,\gamma\mathbf{k}}
}{
\omega^2-\varepsilon_{f,\gamma}(\mathbf{k})^2-D_{f,\gamma\mathbf{k}}^2
}.
\label{eq:hf_anomalous}
\end{align}
These equations show the
poles of $\Sn_{\gamma}$ and $\Sa_{\gamma}$ 
at the same 
particle-hole symmetric energies $\omega=\pm\sqrt{\varepsilon_{f,\gamma}(\mathbf{k})^2+D_{f,\gamma\mathbf{k}}^2}$.

Figure~\ref{fig:hidden_fit} shows the fitting of the 
DCA
self-energies by 
Eqs.~(\ref{eq:hf_normal}) and (\ref{eq:hf_anomalous})
for $\tp=1.0<\tpc$ 
and for $\tp=2.0>\tpc$ 
at $\delta=0.02$.
The low-energy pole structures obtained by DCA are accurately reproduced in both 
$\Sn_{\gamma}$ and $\Sa_{\gamma}$.
This almost perfect fitting at low energies
supports the picture 
that
a quasiparticle 
hybridizes with 
the emergent fermionic degree of freedom $f_{\gamma}$.

The best fitting is obtained at
$(\varepsilon_{f,\gamma}(\mathbf{k}),\,D_{f,\gamma\mathbf{k}},\,|V_{\gamma\mathbf{k}}|)
=(-0.106,\,0.378,\,1.36)$ for $\Mb$ at $\tp=1.0$ and
$(-0.011,\,-0.589,\,1.67)$ for $\Ga$ at $\tp=2.0$. Since
$|\varepsilon_{f,\gamma}|\lesssim0.1$, the corresponding pole energies are
$E_{\gamma}=\sqrt{\varepsilon_{f,\gamma}^2+D_{f,\gamma\mathbf{k}}^2}\simeq|D_{f,\gamma\mathbf{k}}|
=0.39$ and $0.59$, respectively.
These are much smaller than $U=12t$
and the bandwidth $W=8t+2\tp$, 
placing the poles deep inside the
low-energy window. The subdominant
high-energy corrections are detailed in Sec.~S4.

The
hidden-fermion model
explains 
the pole-to-pole cancellation, 
as 
derived
analytically in Sec.~S3 of the Supplemental Material. The quality of the fitting, together with the consistency with this 
analytical result,
strongly supports the conclusion that 
the self-energy poles
originate from a 
hidden-fermion degree of freedom.

Similar hidden-fermion models have
been shown to quantitatively reproduce the 
self-energy of the single-layer repulsive Hubbard model, 
as well as 
of the strong-coupling attractive Hubbard model \cite{Sakai2015PRB,Sakai2016PRB,Sakai2018PRB}, 
where electrons form strongly bound Cooper pairs.
The 
applicability
of the 
hidden-fermion
model in the present bilayer system 
therefore indicates the presence of strongly bound Cooper pairs, which are presumably formed between the layers in this system.
The hidden 
fermion supplies both the 
pseudogap and enhanced superconductivity through the hybridization with low-energy electrons,
acting as a unified mechanism for the two phenomena. 
Specific to the bilayer system, $\tp$ acts as an external knob 
to control the dispersion of the hidden fermion, allowing us to 
optimize
the $s^{\pm}$ pairing. 

\textit{Summary.}
Using DCA+ED, we clarified the real-frequency structure of the self-energy in the bilayer Hubbard model. At half filling, the MI--BI crossover with varying $\tp$ is accompanied by the crossing of the self-energy poles of the $\Ga$ and $\Mb$ components. Upon hole doping, the two different configurations of the self-energy poles produce different band-selective pseudogaps in the normal state: an $\Mb$ ($\Ga$)-selective one for $\tp\lesssim\tpc$ ($\tp\gtrsim\tpc$). In the superconducting state, the self-energy poles stabilize $s^{\pm}$-wave pairing with opposite signs in the two components. The order parameter is maximized near the MI--BI boundary, where the self-energy poles located at low energies reinforce the pairing for both $\Ga$ and $\Mb$.

We have shown that the self-energy structure admits a hidden-fermion
interpretation common to the strongly correlated superconductivity in the single-layer Hubbard model, with a crucial difference specific to the bilayer: $\tp$ acts as an external knob on the pole positions and hence on the pairing strength. The bilayer structure thus offers an unprecedented route to optimizing strongly correlated superconductivity. In systems that realize a single-orbital bilayer Hubbard model near half filling---such as bilayer optical-lattice fermions or suitably engineered bilayer materials---these predictions could be tested directly: ARPES should resolve the band-selective pseudogap that switches between the bonding and antibonding sheets across $\tp\simeq\tpc$, and tuning $\tp$ should trace the
maximization of the pairing near the MI--BI boundary.
A quantitative description of the multiorbital
nickelate La$_3$Ni$_2$O$_7$, which lies away from this regime, is left for future
work.

\textit{Acknowledgments---}
We thank Yusuke Nomura for a useful discussion. This work is
supported by Grant-in-Aid for Scientific Research from
JSPS, KAKENHI Grant No. 25H01252,
No. 24H00190, No. 26K00650, JST K-Program JPMJKP25Z7, RIKEN
TRIP initiative (RIKEN Quantum, Advanced General
Intelligence for Science Program, Many-body Electron
Systems). 

\bibliographystyle{apsrev4-2}
\bibliography{references}

\end{document}


\onecolumngrid
\vspace{1em}
\begin{center}
  {\large\bfseries Supplemental Material for}\\[0.4em]
  {\large\bfseries ``Self-energy pole optimization of superconductivity in the bilayer Hubbard model''}\\[0.8em]
  Taka-Shi Fujiwara, Shiro Sakai, and Ryotaro Arita
\end{center}
\vspace{0.6em}
\twocolumngrid
\setcounter{section}{0}
\setcounter{equation}{0}
\setcounter{figure}{0}
\setcounter{table}{0}
\renewcommand{\thesection}{S\arabic{section}}
\renewcommand{\theequation}{S\arabic{equation}}
\renewcommand{\thefigure}{S\arabic{figure}}
\renewcommand{\thetable}{S\arabic{table}}

\section{Computational details of the DCA scheme}
\label{app:dca}
\subsection{Momentum-space coarse graining}

In the dynamical cluster approximation (DCA) \cite{Hettler1998,Maier2005RMP}, the lattice self-energy is
approximated by a function that is piecewise constant in momentum,
\begin{align}
\Sigma_{\gamma}(\mathbf{k},\omega)\simeq\Sigma_{\gamma}(\mathbf{K},\omega)
\qquad(\mathbf{k}\in P_{\mathbf{K}}),
\label{eq:dca_piecewise}
\end{align}
where $P_{\mathbf{K}}$ denotes the patch surrounding the cluster momentum $\mathbf{K}$.
The Brillouin zone is partitioned into patches of equal area centered at the
cluster momenta, and the lattice degrees of freedom inside each patch are
coarse grained according to Eq.~(2).

\begin{figure}[H]
\centering
\safeincludegraphics[width=0.5\linewidth]{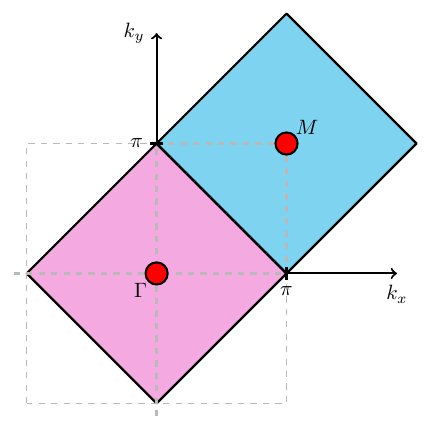}
\caption{Partition of the in-plane Brillouin zone into the two DCA patches used
in this work, centered at the cluster momenta $\mathbf{K}=(0,0)$ and $(\pi,\pi)$.
The patch around $(0,0)$ contains the $\alpha$ (antibonding) Fermi surface and
the patch around $(\pi,\pi)$ contains the $\beta$ (bonding) Fermi surface,
defining the $\Ga$ and $\Mb$ components. Within each patch the self-energy is
approximated as 
$\mathbf{k}$-independent
[Eq.~(\ref{eq:dca_piecewise})].}
\label{fg:dca_patches}
\end{figure}

For the present bilayer calculations, we use the two cluster momenta
$\mathbf{K}=(0,0)$ and $(\pi,\pi)$, so that the in-plane Brillouin zone is divided
into the two patches sketched in Fig.~\ref{fg:dca_patches}: the central region
around $\mathbf{K}=(0,0)$, which contains the $\alpha$-band Fermi surface, and the
corner region around $\mathbf{K}=(\pi,\pi)$, which contains the $\beta$-band Fermi
surface. Together with the two layers, this amounts to $N_c=4$ cluster
orbitals. Because each patch is dominated by a single low-energy band, the
coarse graining directly realizes the band--momentum combined labels $\Ga$ and
$\Mb$ used throughout the main text.

\subsection{Exact-diagonalization (ED) impurity solver}

The cluster problem obtained after the coarse graining is represented by 
the $N_c=4$ interacting cluster
orbitals coupled to $N_b=8$ noninteracting bath sites, and is solved by the
Lanczos method at an artificial inverse temperature $\beta=200$, which
practically yields the ground-state result. 
This method allows the cluster Green's function and self-energy to be evaluated
directly on the real-frequency axis through a continued-fraction 
expansion \cite{Gagliano1987}, without the analytic continuation from the Matsubara-frequency axis required
by quantum Monte Carlo solvers; this is essential for resolving the
low-energy self-energy poles 
of interest here.

In each self-consistency 
cycle,
the bath parameters---the bath levels and
the cluster--bath hybridizations---are 
determined by fitting
the cluster Weiss
field $\mathcal{G}_{0,\gamma}(\mathbf{K},\ii\omega_n)$ with the finite bath.
The bath parameters are optimized by minimizing the
distance function
\begin{align}
d=\sum_{\gamma,n}
\left|
\mathcal{G}_{0,\gamma}(\mathbf{K},\ii\omega_n)
-\mathcal{G}^{\rm ED}_{0,\gamma}(\mathbf{K},\ii\omega_n)
\right|^2,
\label{eq:dca_distance}
\end{align}
evaluated at the fermionic Matsubara frequencies 
$\omega_n=(2n-1)\pi T$,
where $\mathcal{G}^{\rm ED}_{0,\gamma}$ is the Weiss field reproduced by 
the $N_b$ bath sites.

Once the bath parameters are fixed, the discrete cluster impurity Hamiltonian
$H_{\mathrm{cl}}$ is constructed, and its ground state $|\psi_0\rangle$ with
energy $E_0$ is obtained by the Lanczos method. Because the diagonalization is
carried out on the real-space cluster, the cluster Green's function is
evaluated in the real-space--layer basis, in which $c_{il\sigma}$ ($c_{il\sigma}^\dagger$) annihilates (creates)
an electron at cluster site $i$, layer $l$, and spin $\sigma$:
\begin{align}
G^{ll'}_{{\rm cl},\sigma}(i,j;z)
=&\,
\langle\psi_0|\,
c_{il\sigma}\,
\frac{1}{z+E_0-H_{\rm cl}}\,
c_{jl'\sigma}^{\dagger}\,
|\psi_0\rangle
\nonumber\\
&+
\langle\psi_0|\,
c_{jl'\sigma}^{\dagger}\,
\frac{1}{z-E_0+H_{\rm cl}}\,
c_{il\sigma}\,
|\psi_0\rangle .
\label{eq:lanczos_gf}
\end{align}
Each resolvent is generated directly on the real-frequency axis as a
continued fraction from the Lanczos tridiagonalization of $H_{\rm cl}$. Taking
the particle part of a diagonal element ($i=j$, $l=l'$) and starting from the
seed state $|\phi_0\rangle=c_{il\sigma}^{\dagger}|\psi_0\rangle$, the
tridiagonalization of $H_{\rm cl}$ in the Krylov space spanned by
$\{H_{\rm cl}^n|\phi_0\rangle\}_{n=0,1,2,\cdots}$ generates the coefficients $\{a_n,b_n\}$, so that
\begin{align}
&\langle\psi_0|\,
c_{il\sigma}\,
\frac{1}{z+E_0-H_{\rm cl}}\,
c_{il\sigma}^{\dagger}\,
|\psi_0\rangle
\nonumber\\
&\quad=
\cfrac{\langle\phi_0|\phi_0\rangle}
{z+E_0-a_0-\cfrac{b_1^{2}}{z+E_0-a_1-\cfrac{b_2^{2}}{z+E_0-a_2-\cdots}}},
\label{eq:lanczos_cf}
\end{align}
where $a_n=\langle\phi_n|H_{\rm cl}|\phi_n\rangle$ and $b_n^{2}$ are the squared
off-diagonal Lanczos coefficients; the hole part and the off-diagonal
($i\neq j$ or $l\neq l'$) elements are obtained analogously, and the anomalous
components needed in the superconducting state are obtained in a similar way
within the Nambu representation. The real-space--layer Green's function is
finally Fourier transformed over the cluster sites and rotated into the
bonding--antibonding basis 
[$c_{i\sigma}^{\alpha}=(c_{i1\sigma}-c_{i2\sigma})/\sqrt{2}$,
$c_{i\sigma}^{\beta}=(c_{i1\sigma}+c_{i2\sigma})/\sqrt{2}$]
to yield the band--cluster-momentum
components $G_{{\rm cl},\gamma}(\mathbf{K},z)$ used in the main text.

The cluster self-energy is then obtained from the Dyson equation
$\Sigma_{\gamma}(\mathbf{K},\ii\omega_n)
=\mathcal{G}_{0,\gamma}(\mathbf{K},\ii\omega_n)^{-1}
-G_{{\rm cl},\gamma}(\mathbf{K},\ii\omega_n)^{-1}$,
coarse grained through Eq.~(2) to update $\bar{G}_{\gamma}$
and hence the Weiss field
$\mathcal{G}_{0,\gamma}^{-1}=\bar{G}_{\gamma}^{-1}+\Sigma_{\gamma}$, and the loop
is iterated until self-consistency is reached.

\section{Pseudogap spectra at $\dd=0.02$}
\label{app:pseudogap}

\begin{figure*}[t]
\centering
\begin{minipage}[t]{0.49\textwidth}
\centering
\safeincludegraphics[width=1.005\linewidth]{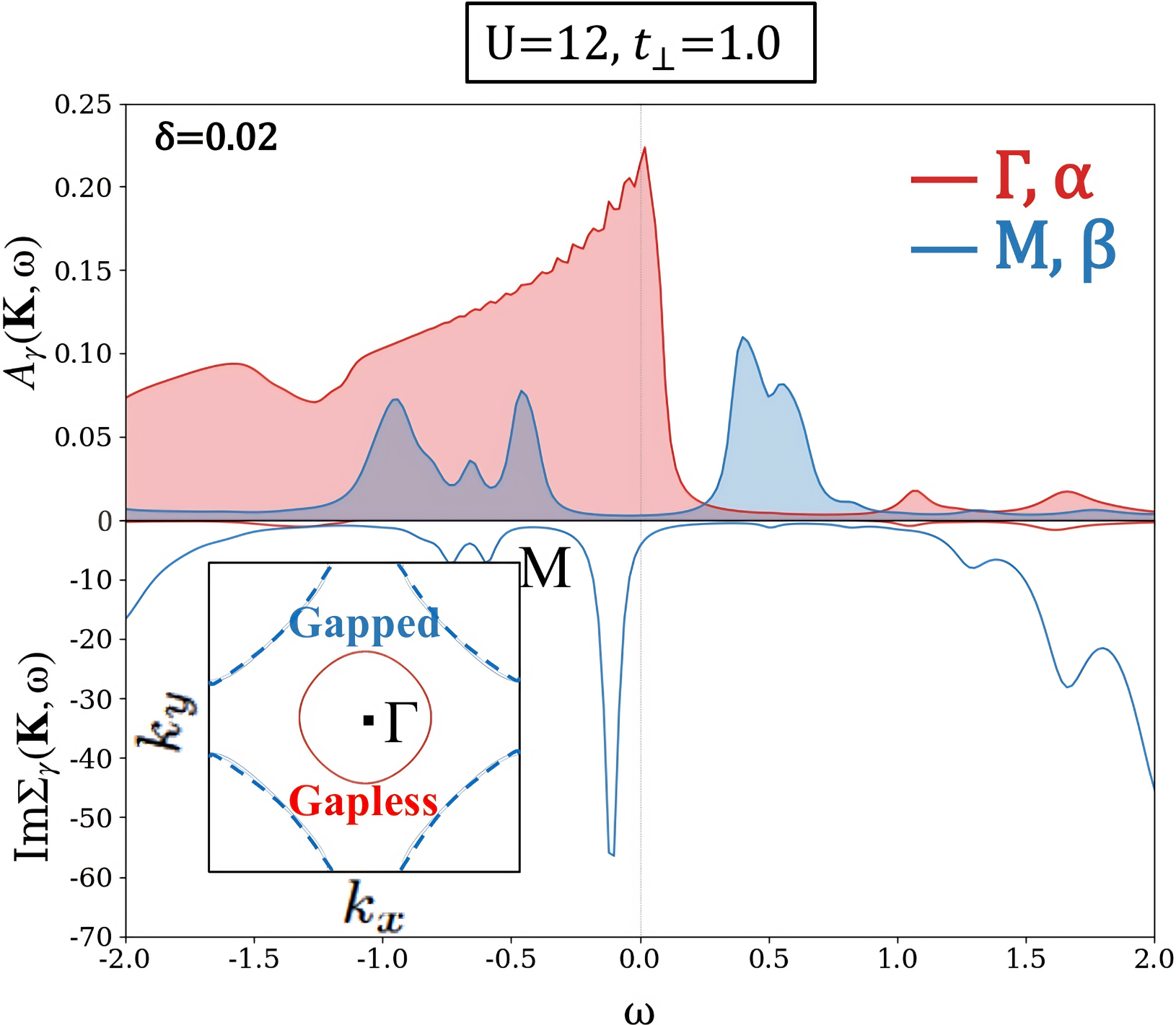}
\end{minipage}\hfill
\begin{minipage}[t]{0.49\textwidth}
\centering
\safeincludegraphics[width=\linewidth]{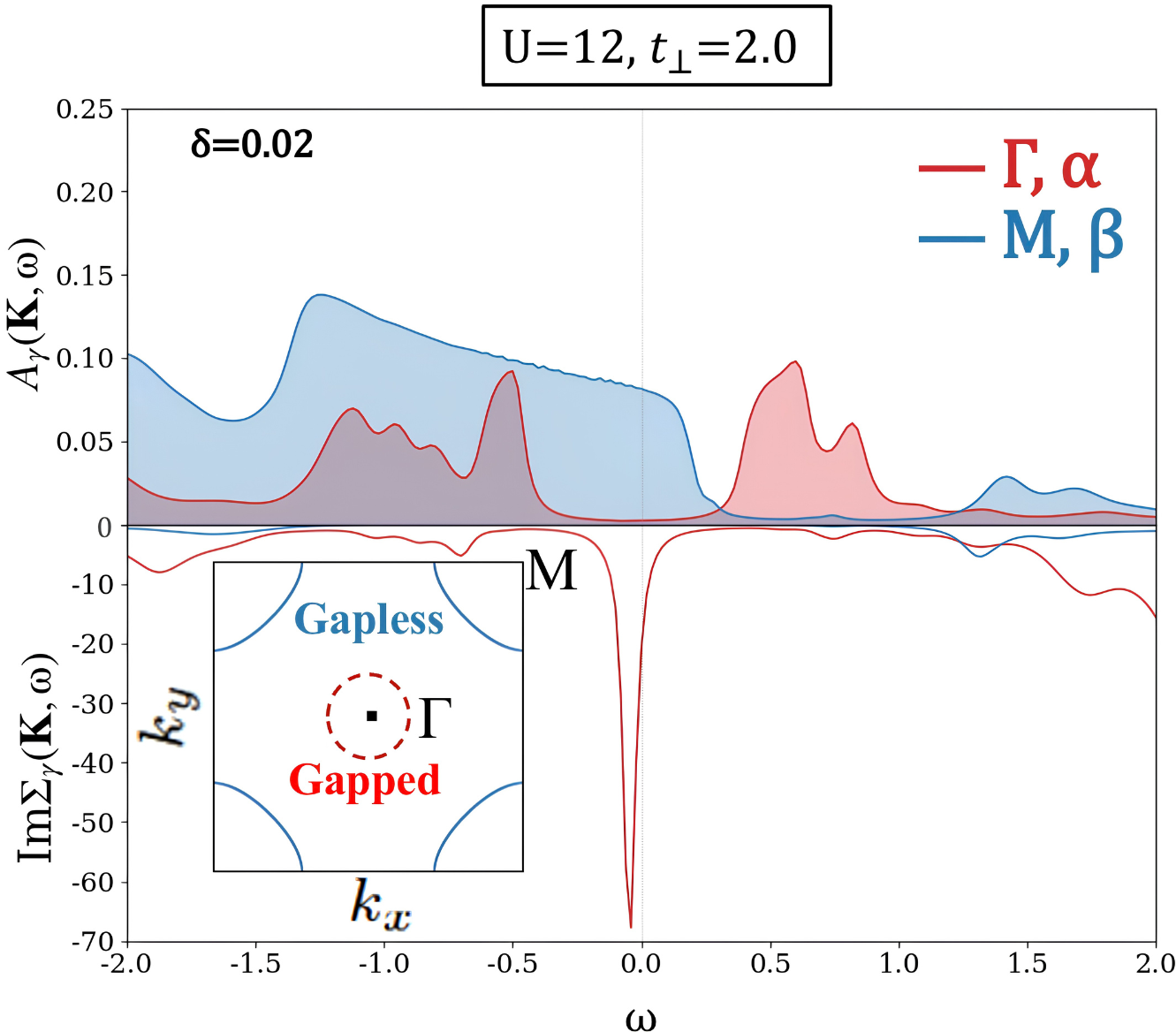}
\end{minipage}
\caption{Spectral function $A_{\gamma}(\mathbf{K},\omega)$ and the imaginary part of the normal self-energy
${\rm Im}\,\Sn_{\gamma}(\mathbf{K},\omega)$ in the normal state at $U=12$ and $\dd=0.02$,
for $\tp=1.0$ (left) and $\tp=2.0$ (right). A pseudogap opens in the $\Mb$ component
for $\tp=1.0$ and in the $\Ga$ component for $\tp=2.0$, while the other
component remains gapless. The insets to the bottom panels schematically illustrate the corresponding
Fermi surfaces, where the gapped (gapless) component is drawn with a dashed (solid) curve.
}
\label{fg:PG}
\end{figure*}

This section presents the detailed spectra underlying the discussion of the
band-selective pseudogap in the main text.
Figure~\ref{fg:PG} shows the spectral functions
$A_{\gamma}(\mathbf{K},\omega)$ and the imaginary part of the normal self-energies ${\rm Im}\,\Sn_{\gamma}(\mathbf{K},\omega)$
obtained in the normal-state calculations at $U=12$ and $\dd=0.02$,
for $\tp=1.0$ (left) and $\tp=2.0$ (right).
On the Mott side ($\tp=1.0$), the self-energy pole remaining near the occupied
band suppresses the low-energy spectral weight of the $\Mb$ component and opens
a pseudogap, while the $\Ga$ component stays gapless.
On the band-insulator side ($\tp=2.0$), the roles of the two components are
interchanged, and the pseudogap opens in the $\Ga$ component.

\section{Pole-to-pole cancellation in the hidden-fermion model}
\label{app:pole_cancel}
In this section, we 
reproduce
the pole-to-pole cancellation observed in the
main text 
in terms of
the hidden-fermion model \cite{Sakai2016PRL}. The component index $\gamma$ and the
momentum $\mathbf{k}$ are suppressed below for brevity, and the fitting parameters are taken to be
real, so that the complex conjugation in Eq.~(4) is immaterial to
the residue analysis. Retaining only the relevant low-energy pole, the normal
and anomalous self-energies [Eqs.~(7) and
(8)] read
\begin{align}
\Sn(\omega)&=s+\frac{V^2(\omega+\varepsilon_f)}{D(\omega)},
\label{eq:pc_Sn}\\
\Sa(\omega)&=D_c-\frac{V^2 D_f}{D(\omega)},
\label{eq:pc_Sa}
\end{align}
with the common denominator
\begin{align}
D(\omega)\equiv\omega^2-\varepsilon_f^2-D_f^2 .
\label{eq:hidden_denom}
\end{align}
Both self-energies have simple poles at the particle-hole symmetric energies
\begin{align}
\omega_{\rm pole}=\pm\sqrt{\varepsilon_f^2+D_f^2}.
\label{eq:pc_pole_pos}
\end{align}
Since $D(\omega)=(\omega-\omega_{\rm pole})(\omega+\omega_{\rm pole})$, one has $D'(\omega_{\rm pole})=2\omega_{\rm pole}$ and
$1/D(\omega)\simeq[2\omega_{\rm pole}(\omega-\omega_{\rm pole})]^{-1}$, so that
the residues of Eqs.~(\ref{eq:pc_Sn}) and (\ref{eq:pc_Sa}) are
\begin{align}
{\rm Res}[\Sn,\omega_{\rm pole}]
&=\frac{V^2(\omega_{\rm pole}+\varepsilon_f)}{2\omega_{\rm pole}}\equiv R_n,
\label{eq:pc_res_Sn}\\
{\rm Res}[\Sa,\omega_{\rm pole}]
&=-\frac{V^2 D_f}{2\omega_{\rm pole}}\equiv R_a .
\label{eq:pc_res_Sa}
\end{align}

We next examine the feedback term $W$ of Eq.~(4),
\begin{align}
W(\omega)&=\frac{[\Sa(\omega)]^2}{\mathcal{D}_W(\omega)},\nonumber\\
\qquad
\mathcal{D}_W(\omega)&\equiv\omega+\varepsilon_c+\Sn(-\omega)^{*} .
\label{eq:pc_W}
\end{align}
Using the evenness $D(-\omega)=D(\omega)$, the reflected normal self-energy reads
\begin{align}
\Sn(-\omega)=s+\frac{V^2(\varepsilon_f-\omega)}{D(\omega)},
\label{eq:pc_Sn_refl}
\end{align}
which carries the \emph{same} poles at $\omega=\omega_{\rm pole}$. Since the
remaining terms of $\mathcal{D}_W$ are regular there, the residue of the
denominator is (real, so the conjugation is irrelevant)
\begin{align}
{\rm Res}[\mathcal{D}_W,\omega_{\rm pole}]
=\frac{V^2(\varepsilon_f-\omega_{\rm pole})}{2\omega_{\rm pole}}\equiv R_d .
\label{eq:pc_res_DW}
\end{align}
Near $\omega=\omega_{\rm pole}$, the numerator of $W$ thus diverges as
$[\Sa]^2\sim R_a^2/(\omega-\omega_{\rm pole})^2$ while the denominator diverges
as $\mathcal{D}_W\sim R_d/(\omega-\omega_{\rm pole})$, so their ratio retains
only a \emph{simple} pole,
\begin{align}
W(\omega)
\simeq
\frac{R_a^2/(\omega-\omega_{\rm pole})^2}{R_d/(\omega-\omega_{\rm pole})}
=\frac{R_a^2/R_d}{\omega-\omega_{\rm pole}}+\mathcal{O}(1),
\end{align}
and hence
\begin{align}
{\rm Res}[W,\omega_{\rm pole}]
=\frac{R_a^2}{R_d}
=\frac{V^2 D_f^2}{2\omega_{\rm pole}(\varepsilon_f-\omega_{\rm pole})} .
\label{eq:pc_res_W}
\end{align}
We note that the regular ($\omega$-finite) parts of $\Sa$ and
$\mathcal{D}_W$---including $\varepsilon_c$, the static shift $s$, and the
high-energy corrections discussed in Sec.~\ref{app:hf_fitting}---contribute only
to the $\mathcal{O}(1)$ background and drop out of the residue, so that
${\rm Res}[W,\omega_{\rm pole}]$ is fixed entirely by the hidden-fermion pole
parameters through $R_a$ and $R_d$.

Adding Eqs.~(\ref{eq:pc_res_Sn}) and (\ref{eq:pc_res_W}), we obtain
\begin{align}
&{\rm Res}[\Sn+W,\omega_{\rm pole}]\nonumber\\
&=\frac{V^2(\omega_{\rm pole}+\varepsilon_f)}{2\omega_{\rm pole}}
+\frac{V^2 D_f^2}{2\omega_{\rm pole}(\varepsilon_f-\omega_{\rm pole})}
\nonumber\\
&=\frac{V^2}{2\omega_{\rm pole}}\,
\frac{(\omega_{\rm pole}+\varepsilon_f)(\varepsilon_f-\omega_{\rm pole})+D_f^2}
{\varepsilon_f-\omega_{\rm pole}}
\nonumber\\
&=\frac{V^2}{2\omega_{\rm pole}}\,
\frac{\varepsilon_f^2-\omega_{\rm pole}^2+D_f^2}{\varepsilon_f-\omega_{\rm pole}}
=0 ,
\label{eq:residue_cancel}
\end{align}
where the numerator in the last line vanishes  
because
$\varepsilon_f^2+D_f^2=\omega_{\rm pole}^2$. Thus, the poles of $\Sn$ and $W$,
originating from the same hidden-fermion degree of freedom, cancel exactly in
the combination $\Sn+W$ that enters the spectral function through
Eq.~(3). This analytically explains the disappearance of the
singular structures in ${\rm Im}(\Sn+W)$ observed numerically in
Fig.~3. The same argument applies independently to the two
poles $\omega_{\rm pole}=\pm\sqrt{\varepsilon_f^2+D_f^2}$ and to each component
$\gamma$.
\section{High-energy corrections and hidden-fermion fitting parameters}
\label{app:hf_fitting}
\subsection{General expression and high-energy corrections}

In general, the self-energy of each component contains, besides the relevant
low-energy pole, additional 
structures
at higher energy.
Because the hidden-fermion model of Eq.~(6) focuses on the low-energy region, the effect of the high-energy structures should be incorporated in the model parameters \cite{Sakai2016PRB,Sakai2023}.
If we represent these high-energy structures with additional hidden fermions labeled by $\nu=2, 3, \dots$, 
we obtain
\begin{align}
\Sn_{\gamma}(\mathbf{k},\omega)&=s_{\gamma\mathbf{k}}+\sum_{\nu}
\frac{V_{\nu,\gamma\mathbf{k}}^2\,(\omega+\varepsilon_{f_\nu,\gamma}(\mathbf{k}))}
{\omega^2-\varepsilon_{f_\nu,\gamma}(\mathbf{k})^2-D_{f_\nu,\gamma\mathbf{k}}^2},
\label{eq:hf_general_nor}\\
\Sa_{\gamma}(\mathbf{k},\omega)&=D_{c,\gamma\mathbf{k}}-\sum_{\nu}
\frac{V_{\nu,\gamma\mathbf{k}}^2\,D_{f_\nu,\gamma\mathbf{k}}}
{\omega^2-\varepsilon_{f_\nu,\gamma}(\mathbf{k})^2-D_{f_\nu,\gamma\mathbf{k}}^2}
\label{eq:hf_general_ano}
\end{align}
in the same way as Eq.~(6).
Here, the $\nu=1$ term is the low-energy pole retained in the main text, for which we
abbreviate $\varepsilon_{f,\gamma}(\mathbf{k})\equiv\varepsilon_{f_1,\gamma}(\mathbf{k})$,
$D_{f,\gamma\mathbf{k}}\equiv D_{f_1,\gamma\mathbf{k}}$, and
$V_{\gamma\mathbf{k}}\equiv V_{1,\gamma\mathbf{k}}$. 
We assume that the other
poles ($\nu\ge2$) lie at high energy,
$E_{\nu,\gamma\mathbf{k}}\equiv\sqrt{\varepsilon_{f_\nu,\gamma}(\mathbf{k})^2+D_{f_\nu,\gamma\mathbf{k}}^2}$,
compared with the low-energy 
region $|\omega|\lesssim E_{1,\gamma\mathbf{k}}$
of interest.
Expanding their contribution up to linear order in $\omega$, 
we obtain
\begin{align}
\sum_{\nu\ge2}
\frac{V_{\nu,\gamma\mathbf{k}}^2\,(\omega+\varepsilon_{f_\nu,\gamma}(\mathbf{k}))}
{\omega^2-E_{\nu,\gamma\mathbf{k}}^2}
&\simeq a_{\gamma\mathbf{k}}-b_{\gamma\mathbf{k}}\,\omega,
\\
-\sum_{\nu\ge2}
\frac{V_{\nu,\gamma\mathbf{k}}^2\,D_{f_\nu,\gamma\mathbf{k}}}
{\omega^2-E_{\nu,\gamma\mathbf{k}}^2}
&\simeq D_{{\rm HE},\gamma\mathbf{k}},
\end{align}
with
\begin{align}
a_{\gamma\mathbf{k}}&=-\sum_{\nu\ge2}
\frac{V_{\nu,\gamma\mathbf{k}}^2\,\varepsilon_{f_\nu,\gamma}(\mathbf{k})}{E_{\nu,\gamma\mathbf{k}}^2},\nonumber\\
b_{\gamma\mathbf{k}}&=\sum_{\nu\ge2}\frac{V_{\nu,\gamma\mathbf{k}}^2}{E_{\nu,\gamma\mathbf{k}}^2},
\nonumber\\
D_{{\rm HE},\gamma\mathbf{k}}&=\sum_{\nu\ge2}
\frac{V_{\nu,\gamma\mathbf{k}}^2\,D_{f_\nu,\gamma\mathbf{k}}}{E_{\nu,\gamma\mathbf{k}}^2}.
\end{align}
Here $a_{\gamma\mathbf{k}}$ is an $\omega$-independent shift, $b_{\gamma\mathbf{k}}$
renormalizes the $c$-fermion band, and $D_{{\rm HE},\gamma\mathbf{k}}$ is a static
anomalous contribution from the high-energy poles. Combining these corrections
with the $\nu=1$ pole, the expressions used for the actual fitting read
\begin{align}
\Sn_{\gamma}(\mathbf{k},\omega)&\simeq
\frac{V_{\gamma\mathbf{k}}^2\,(\omega+\varepsilon_{f,\gamma}(\mathbf{k}))}
{\omega^2-\varepsilon_{f,\gamma}(\mathbf{k})^2-D_{f,\gamma\mathbf{k}}^2}
+\tilde a_{\gamma\mathbf{k}}-b_{\gamma\mathbf{k}}\,\omega,
\label{eq:hf_HE_nor}\\
\Sa_{\gamma}(\mathbf{k},\omega)&\simeq
-\frac{V_{\gamma\mathbf{k}}^2\,D_{f,\gamma\mathbf{k}}}
{\omega^2-\varepsilon_{f,\gamma}(\mathbf{k})^2-D_{f,\gamma\mathbf{k}}^2}
+\tilde D_{{\rm HE},\gamma\mathbf{k}},
\label{eq:hf_HE_ano}
\end{align}
where the static shifts are collected into
$\tilde a_{\gamma\mathbf{k}}\equiv a_{\gamma\mathbf{k}}+s_{\gamma\mathbf{k}}$ and
$\tilde D_{{\rm HE},\gamma\mathbf{k}}\equiv D_{{\rm HE},\gamma\mathbf{k}}+D_{c,\gamma\mathbf{k}}$.
For $\tilde a_{\gamma\mathbf{k}}=b_{\gamma\mathbf{k}}=\tilde D_{{\rm HE},\gamma\mathbf{k}}=0$,
Eqs.~(\ref{eq:hf_HE_nor}) and (\ref{eq:hf_HE_ano}) reduce to the single-pole
expressions of the main text, Eqs.~(7) and
(8).

\subsection{Fitting procedure}
The six parameters $\varepsilon_{f,\gamma}(\mathbf{k})$, $D_{f,\gamma\mathbf{k}}$,
$V_{\gamma\mathbf{k}}$, $\tilde a_{\gamma\mathbf{k}}$, $b_{\gamma\mathbf{k}}$, and
$\tilde D_{{\rm HE},\gamma\mathbf{k}}$ of each component are determined by a
least-squares fit of the DCA+ED self-energies, using the same broadening as in
the DCA+ED calculation, $\eta(\omega)=\eta_0\,\min(10,\,1+\omega^{2})$ with
$\eta_0=3\times10^{-2}$, which enters through the complex frequency
$\omega\to\omega+\ii\,\eta(\omega)$. The fit is restricted to the low-energy
window in which the $\nu=1$ pole is well separated from the higher-energy
structures; outside this window additional poles enter the relevant range and
the single-pole description breaks down. The fits shown in
Fig.~4 of the main text are obtained in this way.

For the two fits shown in Fig.~4 of the main text, the
high-energy correction parameters
$(\tilde a_{\gamma\mathbf{k}},\,b_{\gamma\mathbf{k}},\,\tilde D_{{\rm HE},\gamma\mathbf{k}})$
are those listed in Table~\ref{tab:hf_params_tp}, namely,
$(2.10,\,7.04,\,1.41)$ for the $\Mb$ component at $\tp=1.0$ and
$(1.83,\,3.93,\,-0.33)$ for the $\Ga$ component at $\tp=2.0$.

\subsection{$t_\perp$ dependence of the hidden-fermion fit}

\begin{figure}[H]
\centering
\safeincludegraphics[width=\linewidth]{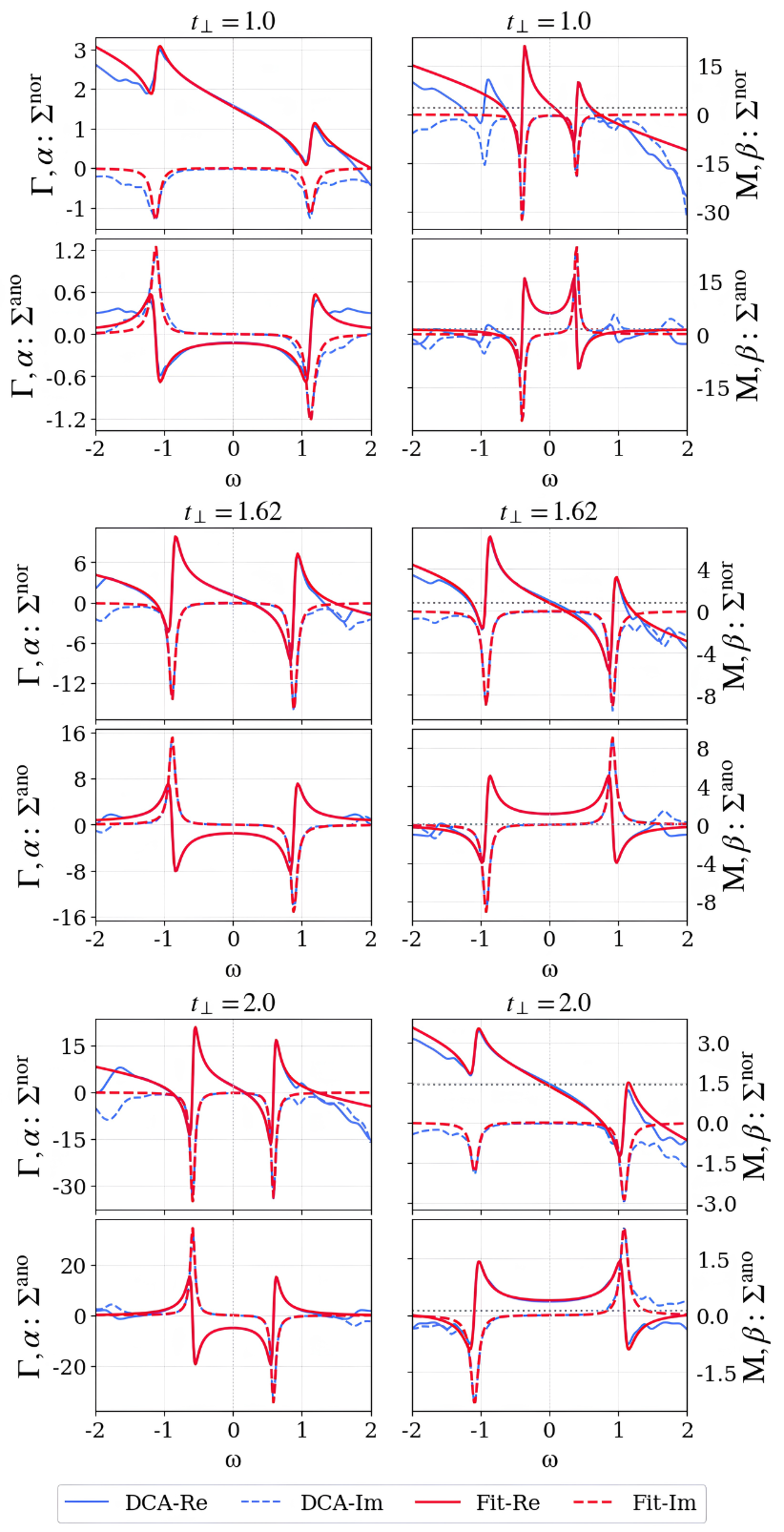}
\caption{Hidden-fermion fits of the normal ($\Sn_{\gamma}$) and anomalous
($\Sa_{\gamma}$) self-energies for the three representative interlayer hoppings
$\tp=1.0$, $1.62$, and $2.0$ (top, middle, and bottom blocks) at $U=12$ and
$\dd=0.02$, straddling the pole crossing at half filling. Within each block the
left (right) column shows the $\Ga$ ($\Mb$) component, and the upper (lower) row
shows $\Sn_{\gamma}$ ($\Sa_{\gamma}$). Blue curves are the DCA 
results and red
curves the hidden-fermion fit [Eqs.~(\ref{eq:hf_HE_nor}) and
(\ref{eq:hf_HE_ano})]; solid (dashed) 
curves
denote the real (imaginary) part.
The low-energy 
pole-like structures are reproduced for both components and both 
channels at every $\tp$. The fitting parameters are listed in Table~\ref{tab:hf_params_tp}.
}
\label{fg:hf_params}
\end{figure}

Figure~\ref{fg:hf_params} demonstrates that the hidden-fermion model reproduces
the DCA 
self-energies over the entire range of $\tp$, not only at a single
value. 

We show $\Sn_{\gamma}$ and $\Sa_{\gamma}$ for both components
$\gamma\in\{\alpha,\beta\}$ for $U=12$ and $\dd=0.02$ at the three representative interlayer hoppings
$\tp=1.0$, $1.62$, and $2.0$, which straddle the pole crossing at half filling.
For all
$\tp$, the fit
[Eqs.~(\ref{eq:hf_HE_nor}) and (\ref{eq:hf_HE_ano})] accurately captures the
low-energy 
pole-like
structures of both the real and imaginary parts of
$\Sn_{\gamma}$ and $\Sa_{\gamma}$ in the $\Ga$ and $\Mb$ components alike. This
systematic agreement, irrespective of $\tp$ and the component, shows that a
single low-energy hidden-fermion pole per component suffices to describe the
self-energy throughout the MI--BI crossover, and supports the picture that a
quasiparticle in each component hybridizes with the emergent fermion
$f_{\gamma}$ across the whole parameter range.

The fitted pole position, read off from the pole energy
$\pm\sqrt{\varepsilon_{f,\gamma}(\mathbf{k})^2+D_{f,\gamma\mathbf{k}}^2}$ of each
component, follows the pole motion discussed in the main text: on the MI side
($\tp=1.0$) the $\Mb$ pole lies closest to $\omega=0$, both poles approach low
energy at $\tp=1.62$, and on the BI side ($\tp=2.0$) the $\Ga$ pole takes over
the low-energy region.
\begin{figure}[t]
\centering
\safeincludegraphics[width=1.01\linewidth]{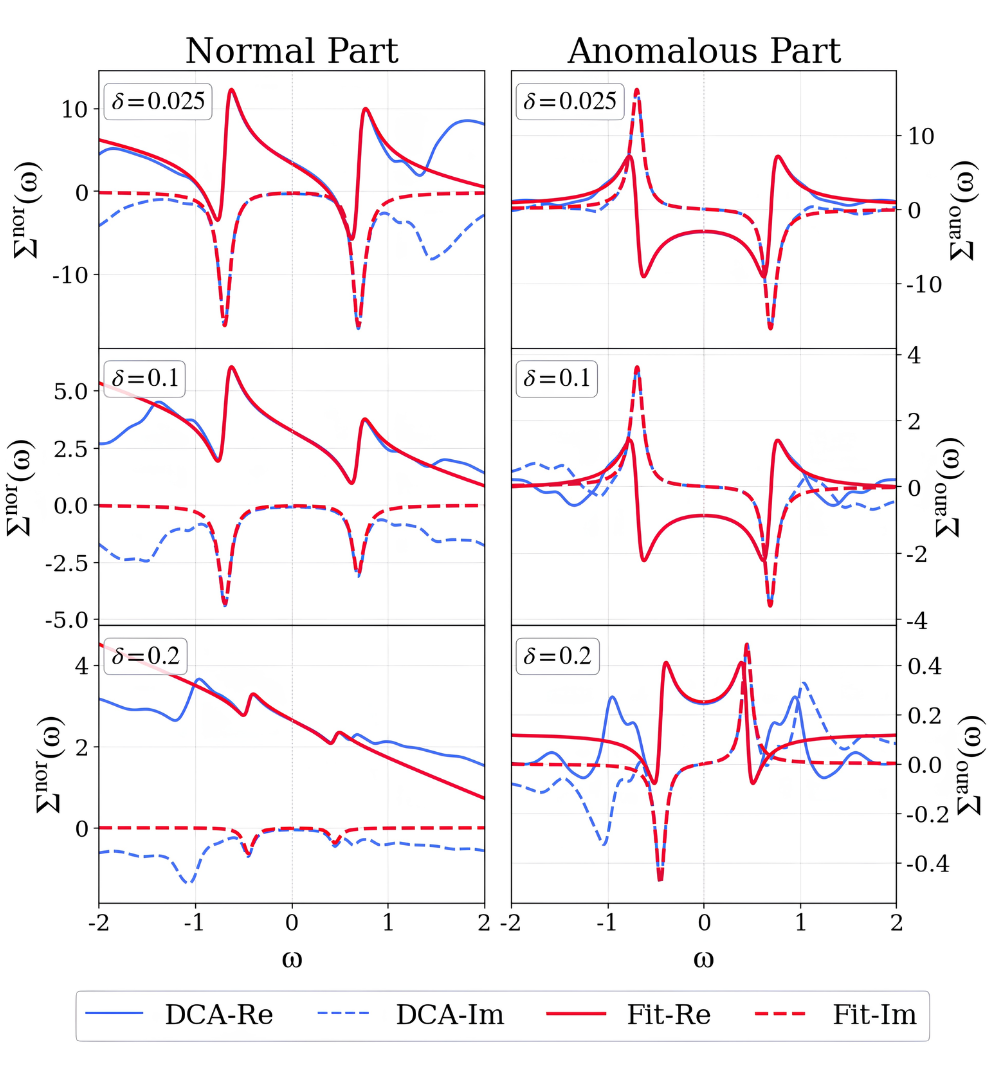}
\caption{Doping dependence of the hidden-fermion fit of the normal
($\Sn_{\gamma}$, left column) and anomalous ($\Sa_{\gamma}$, right column)
self-energies of the $\Ga$ component at $U=8$ and $\tp=2.0$, for $\dd=0.025$,
$0.1$, and $0.2$ (top to bottom). Blue curves are the DCA results and red curves
the hidden-fermion fit [Eqs.~(\ref{eq:hf_HE_nor}) and (\ref{eq:hf_HE_ano})];
solid (dashed) curves denote the real (imaginary) part. The fitting parameters are listed in Table~\ref{tab:hf_params_doping}.
Note that the vertical
scale differs between panels. With increasing doping, the pole structure
associated with the hidden-fermion excitation weakens and, in the large-doping regime
($\dd=0.2$), becomes buried among other structures of the self-energy.}
\label{fg:hf_doping}
\end{figure}
\subsection{Doping dependence of the hidden-fermion fit}
\label{ssec:hf_doping}

The hidden-fermion description is rooted in the proximity to the Mott insulator,
and is therefore expected to be most accurate at small doping and to deteriorate as 
the doping increases.
To examine
this trend, Fig.~\ref{fg:hf_doping} shows $\Sn_{\gamma}$ and $\Sa_{\gamma}$
together with their hidden-fermion fits for the $\Ga$ component at $U=8$ and
$\tp=2.0$, for three hole concentrations $\dd=0.025$, $0.1$, and $0.2$.

At the smallest doping $\dd=0.025$, both $\Sn_{\gamma}$ and $\Sa_{\gamma}$ exhibit
pole structures at the particle-hole symmetric pole energies
$\omega\simeq\pm0.7$, which the single hidden-fermion pole
[Eqs.~(\ref{eq:hf_HE_nor}) and (\ref{eq:hf_HE_ano})] reproduces almost
quantitatively. With increasing doping, the pole feature systematically weakens.
Since the vertical scale differs between panels, the comparable-looking line shapes, in fact, correspond to a rapid decrease in the absolute magnitude of the pole.

For a large doping ($\dd=0.2$),
the low-energy pole is no longer the dominant
feature of the self-energy. The real part of $\Sn_{\gamma}$ is governed by the
smooth, nearly linear background generated by the higher-energy 
structures
(the
$-b_{\gamma\mathbf{k}}\,\omega$ term), on top of which the genuine low-energy pole
appears only as a weak wiggle. The DCA data develop additional structures
away from the pole---a broad background in ${\rm Im}\,\Sn_{\gamma}$ and
secondary features around $\omega\simeq\pm1$ in $\Sa_{\gamma}$ of magnitude
comparable to the central pole---that lie outside the single-pole ansatz and are not captured by the fit. The clean separation of the $\nu=1$ pole from the
remaining structures, on which the single-pole fit relies, is thus gradually lost, and the hidden-fermion pole becomes buried among the other contributions.

\begin{table*}[t]
\caption{Hidden-fermion parameters in Eqs.~(\ref{eq:hf_HE_nor}) and (\ref{eq:hf_HE_ano}), used in fitting Fig.~\ref{fg:hf_params} for the $\Ga$ and $\Mb$ components at $U=12$ and $\dd=0.02$. Since only $V_{\gamma\mathbf{k}}^{2}$ enters Eqs.~(\ref{eq:hf_HE_nor}) and (\ref{eq:hf_HE_ano}), we quote $|V_{\gamma\mathbf{k}}|$. The last column lists the self-energy pole energy $E_{\gamma\mathbf{k}}=\sqrt{\varepsilon_{f,\gamma}(\mathbf{k})^{2}+D_{f,\gamma\mathbf{k}}^{2}}$.}
\label{tab:hf_params_tp}
\footnotesize
\setlength{\tabcolsep}{3pt}
\begin{ruledtabular}
\begin{tabular}{llccccccc}
 & & $|V_{\gamma\mathbf{k}}|$ & $\varepsilon_{f,\gamma}(\mathbf{k})$ & $D_{f,\gamma\mathbf{k}}$ & $\tilde D_{{\rm HE},\gamma\mathbf{k}}$ & $\tilde a_{\gamma\mathbf{k}}$ & $b_{\gamma\mathbf{k}}$ & $E_{\gamma\mathbf{k}}$\\
\hline
$\tp=1.0$ & $\Ga$ & $4.1218\times10^{-1}$ & $-6.8086\times10^{-2}$ & $-1.1259\times10^{0}$ & $2.1871\times10^{-2}$ & $1.5369\times10^{0}$ & $8.2592\times10^{-1}$ & $1.1279\times10^{0}$\\
 & $\Mb$ & $1.3615\times10^{0}$ & $-1.0587\times10^{-1}$ & $3.7817\times10^{-1}$ & $1.4129\times10^{0}$ & $2.1043\times10^{0}$ & $7.0389\times10^{0}$ & $3.9271\times10^{-1}$\\[3pt]
$\tp=1.62$ & $\Ga$ & $1.2785\times10^{0}$ & $4.6942\times10^{-2}$ & $-8.8509\times10^{-1}$ & $3.4617\times10^{-1}$ & $1.1346\times10^{0}$ & $1.9902\times10^{0}$ & $8.8634\times10^{-1}$\\
 & $\Mb$ & $1.0025\times10^{0}$ & $3.1989\times10^{-3}$ & $9.1959\times10^{-1}$ & $3.7875\times10^{-2}$ & $7.6208\times10^{-1}$ & $2.1400\times10^{0}$ & $9.1959\times10^{-1}$\\[3pt]
$\tp=2.0$ & $\Ga$ & $1.6679\times10^{0}$ & $-1.1422\times10^{-2}$ & $-5.8873\times10^{-1}$ & $-3.2640\times10^{-1}$ & $1.8274\times10^{0}$ & $3.9301\times10^{0}$ & $5.8884\times10^{-1}$\\
 & $\Mb$ & $5.5842\times10^{-1}$ & $2.3868\times10^{-1}$ & $1.0625\times10^{0}$ & $1.1394\times10^{-1}$ & $1.4322\times10^{0}$ & $1.1655\times10^{0}$ & $1.0890\times10^{0}$\\
\end{tabular}
\end{ruledtabular}
\end{table*}
 
\begin{table*}[t]
\caption{Hidden-fermion parameters in Eqs.~(\ref{eq:hf_HE_nor}) and (\ref{eq:hf_HE_ano}) used in fitting Fig.~\ref{fg:hf_doping} for the $\Ga$ component at $U=8$, $\tp=2.0$, and the hole concentrations $\dd=0.025$, $0.1$, and $0.2$. Notations are the same as those in Table~\ref{tab:hf_params_tp}.}
\label{tab:hf_params_doping}
\footnotesize
\setlength{\tabcolsep}{3pt}
\begin{ruledtabular}
\begin{tabular}{lccccccc}
 & $|V_{\gamma\mathbf{k}}|$ & $\varepsilon_{f,\gamma}(\mathbf{k})$ & $D_{f,\gamma\mathbf{k}}$ & $\tilde D_{{\rm HE},\gamma\mathbf{k}}$ & $\tilde a_{\gamma\mathbf{k}}$ & $b_{\gamma\mathbf{k}}$ & $E_{\gamma\mathbf{k}}$\\
\hline
$\dd=0.025$ & $1.5502\times10^{0}$ & $7.0661\times10^{-5}$ & $-6.9657\times10^{-1}$ & $4.3069\times10^{-1}$ & $3.3535\times10^{0}$ & $2.0836\times10^{0}$ & $6.9657\times10^{-1}$\\
$\dd=0.1$ & $7.3754\times10^{-1}$ & $-1.2361\times10^{-1}$ & $-6.8241\times10^{-1}$ & $-1.0574\times10^{-1}$ & $3.0945\times10^{0}$ & $1.2741\times10^{0}$ & $6.9351\times10^{-1}$\\
$\dd=0.2$ & $2.4583\times10^{-1}$ & $-1.1717\times10^{-1}$ & $4.3543\times10^{-1}$ & $1.2193\times10^{-1}$ & $2.6172\times10^{0}$ & $9.6080\times10^{-1}$ & $4.5092\times10^{-1}$\\
\end{tabular}
\end{ruledtabular}
\end{table*}

\section{Determination of the MI--BI boundary}
\label{app:mi_bi}

In this section, we describe how the phase boundaries at half filling,
superimposed in Fig.~2, are determined.
In the insulating phase, we estimate the size of the excitation gap from the
density of states $A_{\mathrm{tot}}(\omega)=\sum_{\gamma,\mathbf{K}}A_{\gamma}(\mathbf{K},\omega)$ at half filling. To locate the gap
edges, 
we compute the frequency derivative of the
spectral function and adopt the local maximum of
$|dA_{\mathrm{tot}}(\omega)/d\omega|$ closest to $\omega=0$ as the gap edge, i.e.,
the energy at which the spectral weight rises most steeply out of the gap
region. The excitation gap $\Delta$ is then evaluated 
as the energy difference between
the gap edges
determined in this way.

\begin{figure}[t]
\centering
\safeincludegraphics[width=\linewidth]{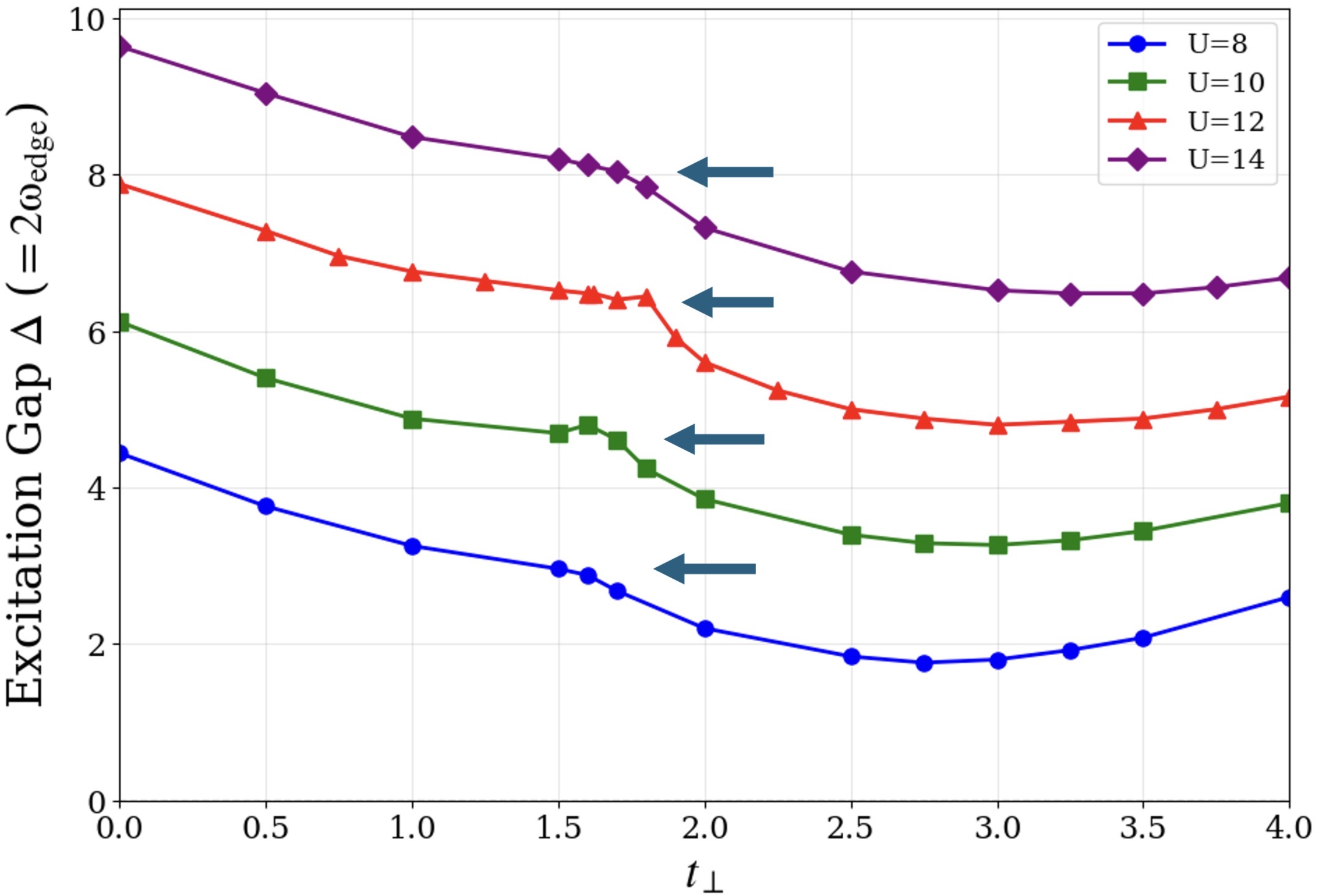}
\caption{Excitation gap $\Delta$ in the half-filling insulating regime as a function of $\tp$,
extracted from the spectral functions calculated for $U=8$, $10$, $12$, and
$14$ with varying $\tp$. The gap edges are determined from the local maxima
of $|dA_{\mathrm{tot}}(\omega)/d\omega|$ near $\omega=0$. The kink in
$\Delta(\tp)$ (indicated by arrows) marks the MI--BI boundary.}
\label{fg:Ex_Gap}
\end{figure}

\begin{figure}[t]
\centering
\safeincludegraphics[width=\linewidth]{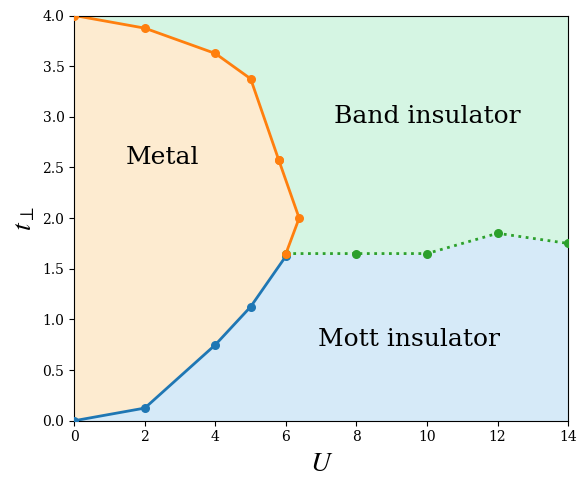}
\caption{Phase diagram of the bilayer Hubbard model at half filling in the
$(U,\tp)$ plane, determined from $\Delta(\tp)$.  The MI--BI boundary,
defined by the kink in $\Delta(\tp)$, nearly coincides with the crossing
point of the self-energy poles shown in Fig.~1.}
\label{fg:phase_diagram}
\end{figure}

Figure~\ref{fg:Ex_Gap} shows $\Delta$ as a function of $\tp$ for several
values of $U$ in the insulating regime.
For each $U$, $\Delta(\tp)$ exhibits a kink at a certain value
of $\tp\simeq\tpc$, reflecting the change in the character of the insulating gap. 
We identify this kink with the MI--BI boundary. This criterion is slightly different from that of Ref.~\cite{KancharlaOkamoto2007}, where the MI--BI crossover at large $U$ was identified with the point at which the charge gap turns from a decrease to an increase upon increasing $\tp$. In the charge-gap data of Ref.~\cite{KancharlaOkamoto2007}, however, this turning point appears as a kink of $\Delta_c(\tp)$, i.e., a sharp minimum (at $U=12t$) or a bend into a nearly flat behavior (at $U=16t$); identifying the crossover with this feature corresponds to the boundary defined here, and the two ways of determining the boundary are therefore consistent with each other. 
Notably, the position of the
kink almost coincides with the crossing point of the self-energy poles of the
$\Ga$ and $\Mb$ components discussed in the main text
[e.g., $\tp\simeq 1.62$ at $U=12$; see Fig.~1],
corroborating that the MI--BI crossover is governed by the rearrangement of
the low-energy self-energy poles. The phase diagram thus obtained is shown in
Fig.~\ref{fg:phase_diagram}.

\section{Doping dependence of the superconducting order parameter}
\label{app:op_doping}

\begin{figure}[H]
\centering
\safeincludegraphics[width=1.02\linewidth]{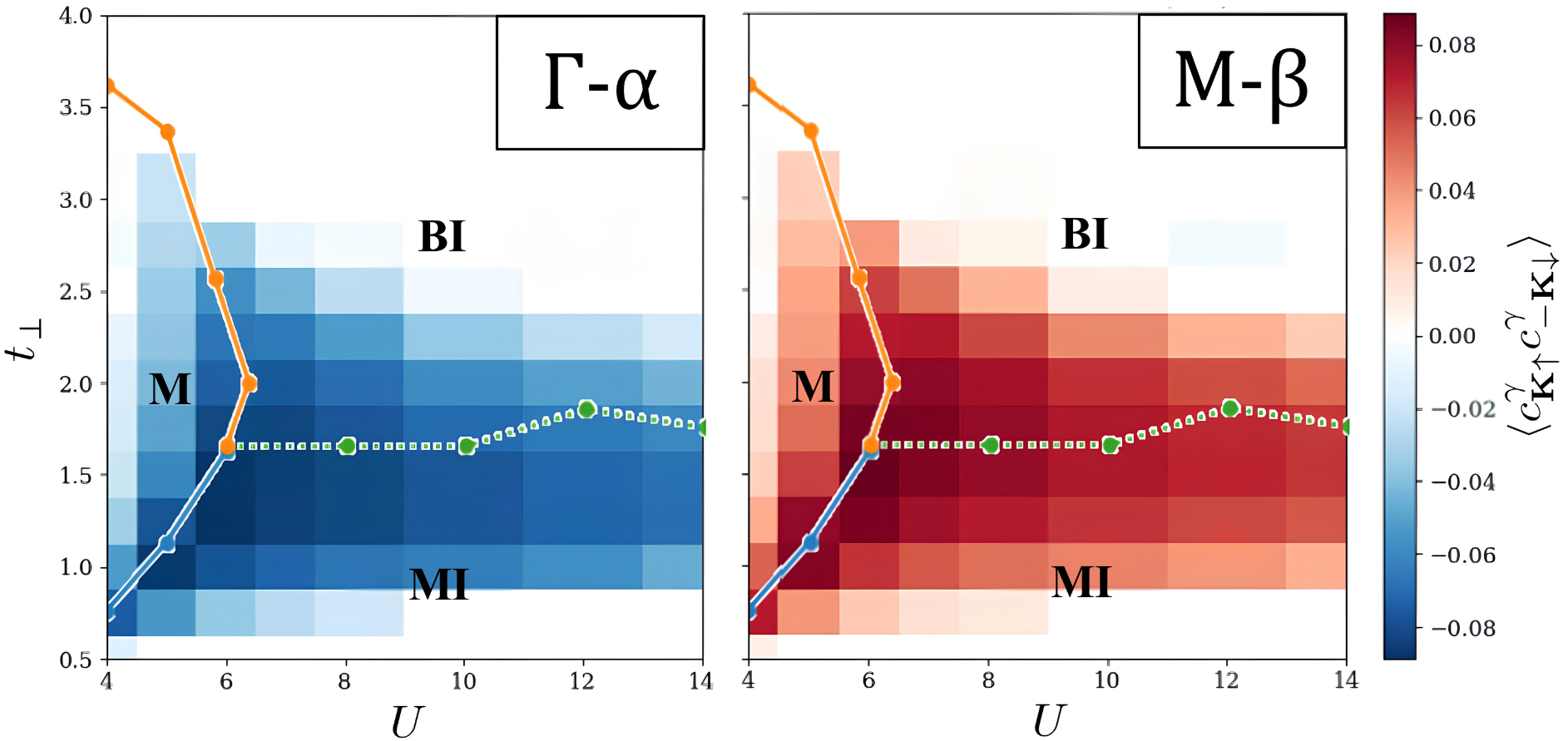}
\caption{Superconducting order parameters in the $(U,\tp)$ plane at $\dd=0.05$
for the $\Ga$ component
[$\langle c_{\Gamma\uparrow}^{\alpha}c_{\Gamma\downarrow}^{\alpha}\rangle$, left]
and the $\Mb$ component
[$\langle c_{\mathrm{M}\uparrow}^{\beta}c_{\mathrm{M}\downarrow}^{\beta}\rangle$, right], on a
common color scale; the opposite signs ($\Ga<0$: blue, $\Mb>0$: red) indicate
$s^{\pm}$-wave symmetry. The half-filling phase diagram (M: metal, MI: Mott
insulator, BI: band insulator) and its boundaries (from Sec.~\ref{app:mi_bi}),
including the MI--BI boundary (green dashed curve), are superimposed. Compared with
$\dd=0.02$ (Fig.~2), the region of large order parameter extends
along the MI--BI boundary toward the strongly correlated (large-$U$) side.}
\label{fg:op_d005}
\end{figure}

In the main text, the superconducting order parameters are shown at $\dd=0.02$
[Fig.~2], where both $\Ga$ and $\Mb$ components are largest near the 
crossing point of the metal--insulator boundary and 
the MI--BI boundary. To examine how this distribution evolves with doping,
Fig.~\ref{fg:op_d005} shows the corresponding $(U,\tp)$ maps of the $\Ga$ and
$\Mb$ order parameters at a larger doping $\dd=0.05$, with the half-filling
phase diagram (Sec.~\ref{app:mi_bi}) superimposed.

The order parameters retain the $s^{\pm}$ character found at $\dd=0.02$: the $\Ga$
and $\Mb$ components carry opposite signs
[$\langle c_{\Gamma\uparrow}^{\alpha}c_{\Gamma\downarrow}^{\alpha}\rangle<0$ and
$\langle c_{\mathrm{M}\uparrow}^{\beta}c_{\mathrm{M}\downarrow}^{\beta}\rangle>0$]. Moreover, the
region where the superconductivity emerges is almost unchanged from
$\dd=0.02$: both components are concentrated in a horizontal stripe centered
on the MI--BI boundary, which nearly coincides with the pole crossing
$\tp\simeq\tpc$ discussed in the main text.

The main difference from $\dd=0.02$ thus lies in the distribution of the
order-parameter magnitude within this stripe: while the largest values are
still found near the 
metal--insulator boundary,
sizable values now extend
along the MI--BI boundary toward 
larger $U$ values.
This behavior can be
interpreted in terms of the doping dependence of the hidden-fermion pole
(see Fig.~\ref{fg:hf_doping} and Table~\ref{tab:hf_params_doping}): with
increasing doping, the intensity of the low-energy self-energy poles
decreases, so that the dynamical pairing enhancement driven by these poles
weakens and the pairing approaches a more conventional, BCS-like static one.
The weakened enhancement around the metal--insulator boundary results in relatively large values of the order parameters at large $U$. 
This trend may also be regarded as reflecting the
doping-induced crossover from a BEC-like to a BCS-like superconducting
state reported for the bilayer Hubbard model in Ref.~\cite{Nomura2025}.

\bibliographystyle{apsrev4-2}
\bibliography{references}